\documentclass[prx,showpacs,twocolumn,superscriptaddress,aps]{revtex4-2}

\usepackage[utf8]{inputenc}
\usepackage[american,british]{babel}
\usepackage[T1]{fontenc}
\usepackage[pdftex]{graphicx}  
\usepackage{graphicx, xcolor}
\usepackage{dcolumn}
\usepackage{bm}
\usepackage{amsmath,amsthm,amssymb}
\usepackage{color}
\usepackage{verbatim}
\usepackage{hyperref}

\usepackage{ulem} % cancel text

\def\ave#1{\langle #1\rangle}

\newcommand{\bra}[1]{\langle #1|}
\newcommand{\ket}[1]{|#1\rangle}

\newcommand{\be}{\begin{equation}}
\newcommand{\ee}{\end{equation}}

\renewcommand{\aa}[1]{{\color{orange} #1}}

\usepackage{bm}% bold math

\begin{document}

\title{Intrinsic dimension of path integrals:\\ data mining quantum criticality and emergent simplicity}

\author{T. Mendes-Santos*}
\affiliation{The Abdus Salam International Centre for Theoretical Physics, strada Costiera 11, 34151 Trieste, Italy}
\affiliation{Max-Planck-Institut für Physik komplexer Systeme, 01187 Dresden, Germany}
\author{A. Angelone*}
\affiliation{The Abdus Salam International Centre for Theoretical Physics, strada Costiera 11, 34151 Trieste, Italy}
\affiliation{SISSA, via Bonomea, 265, 34136 Trieste, Italy}
\author{Alex Rodriguez}
\affiliation{The Abdus Salam International Centre for Theoretical Physics, strada Costiera 11, 34151 Trieste, Italy}
\author{R. Fazio}
\affiliation{The Abdus Salam International Centre for Theoretical Physics, strada Costiera 11, 34151 Trieste, Italy}
\affiliation{Dipartimento di Fisica, Universit\`a di Napoli Federico II, Monte S. Angelo, I-80126 Napoli, Italy}
\author{M. Dalmonte}
\affiliation{The Abdus Salam International Centre for Theoretical Physics, strada Costiera 11, 34151 Trieste, Italy}
\affiliation{SISSA, via Bonomea, 265, 34136 Trieste, Italy}

\begin{abstract}
Quantum many-body systems are characterized by patterns of correlations
  defining highly non-trivial manifolds when interpreted as data structures.
  Physical properties of phases and phase transitions are typically retrieved
  via correlation functions, that are related to observable response functions.
  Recent experiments have demonstrated capabilities to fully characterize
  quantum many-body systems via wave-function snapshots, opening new
  possibilities to analyze quantum phenomena. Here, we introduce a method to
  data mine the correlation structure of quantum partition functions via their
  path integral (or equivalently, stochastic series expansion) manifold. We
  characterize path integral manifolds generated via state-of-the-art Quantum
  Monte Carlo methods utilizing the intrinsic dimension (ID) and the variance
  of distances between nearest neighbors configurations (NN): the former is
  related to data set complexity, while the latter is able to diagnose
  connectivity features of points in configuration space. We show how these
  properties feature universal patterns in the vicinity of quantum criticality,
  that reveal how data structures {\it simplify} systematically at quantum
  phase transitions. This is further reflected by the fact that both ID and
  variance of NN distances exhibit universal scaling behavior in the vicinity
  of second-order and Berezinskii-Kosterlitz-Thouless critical points. Finally,
  we show how non-Abelian symmetries dramatically influence quantum data sets,
  due to the nature of (non-commuting) conserved charges in the quantum case.
  Complementary to neural network representations, our approach represents a
  first elementary step towards a systematic characterization of path integral
  manifolds before any dimensional reduction is taken, that is informative
  about universal behavior and complexity, and can find immediate application
  to both experiments and Monte Carlo simulations.
\end{abstract}

\maketitle
\section{Introduction} 

The path integral (PI) formulation of quantum partition functions is arguably
one of the most basic concepts in quantum many-body theory~\cite{AltlandBook,
Montvay1994, sachdev_2000}. It provides key and generic information about a
given quantum state, typically interpreted via low-order correlation functions.
More refined properties, such as the degree of quantum correlations captured by
entanglement, can also be extracted from path integrals leveraging on advanced
techniques such as the replica trick~\cite{calabrese2009entanglement,
Amico2008}, or by analyzing the topological structure of their degrees of
freedom~\cite{Lingua2021}. On a more general, yet abstract ground, the path
integral of a many-body problem can be construed as a very complex
multi-dimensional manifold embedded in a space that describes both spatial and
imaginary-time coordinates: within this geometrical interpretation, while
low-order properties of the PI manifold (and their relation to physical
observables) are in general well-understood, it is currently unclear if global
properties of such data structures can be informative about physical phenomena
at all, or if they are constrained by universal properties of the many-body
dynamics~\cite{sachdev_2000}.

\begin{figure*}[t]
\begin{center}
{\centering\resizebox*{18.6cm}{!}{\includegraphics*{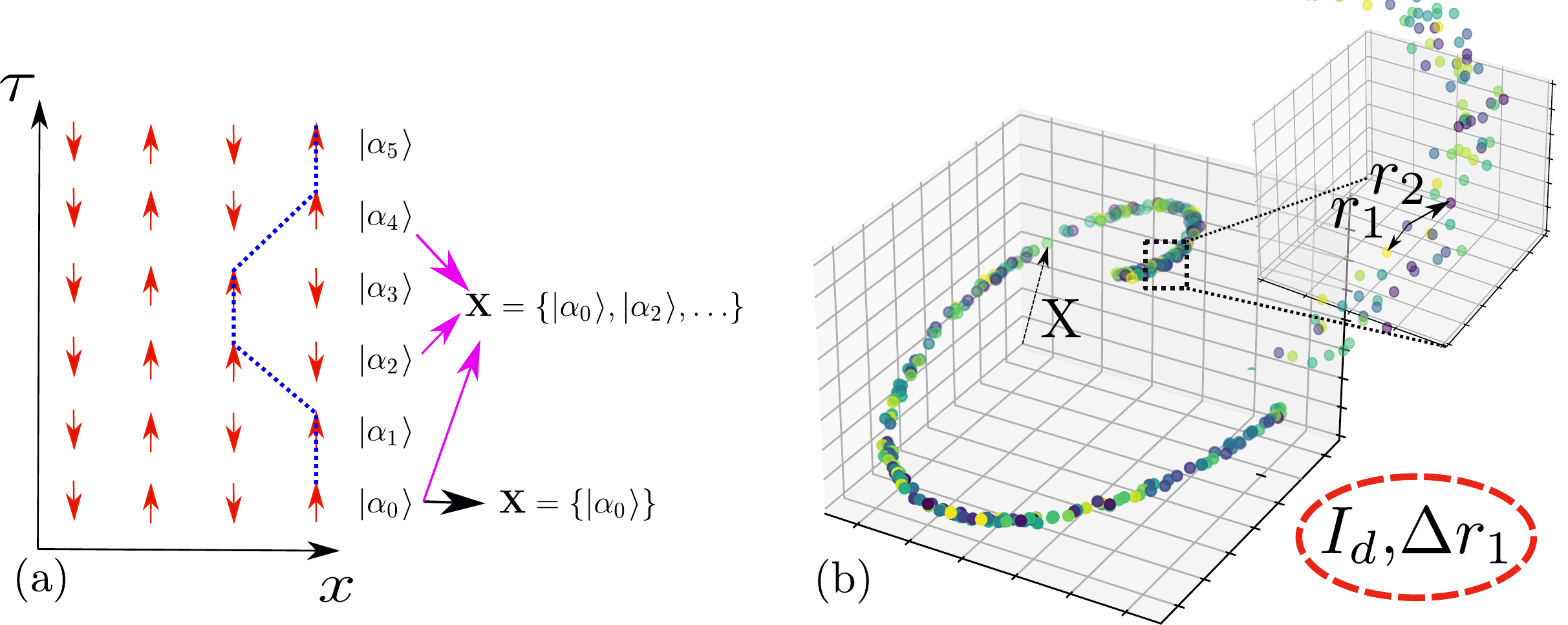}}}
\end{center}
\caption{\textit{Data structure of quantum partition functions and generic data
  set features.} The partition function of a quantum system can be described by
  an extended configuration space with one extra dimension $\tau$ (known as
  \textit{imaginary time}). As an example, we illustrate in panel (a) a
  specific space-time configuration of a system with four spins. Each slice
  $\ket{\alpha_i}$ is defined in terms of the spatial degrees of freedom, i.e.,
  $\ket{\alpha_i} = (S_{1,i}^z,...,S_{N_s,i}^z)$. Here, we consider data sets
  generated by either (i) single or (ii) multiple slices. The data structure of
  quantum partition functions is described by the manifold generated by the set
  of points ${\bf X}$ in a high-dimensional space. To pictorially explain our
  approach, we consider in panel (b) a synthetic data set embedded in three
  dimensions. We investigate generic features of data sets associated to the
  statistics of nearest- and next-nearest-neighbor configuration distances,
  $r_1$ and $r_2$, respectively: namely, the intrinsic dimension, $I_d$, and
  the variance $\Delta r_1$ of the distribution function $f(r_1)$ (see text).
  Our main result is to show that the $I_d$ and $\Delta r_1$ exhibit universal
  critical behavior in the vicinity of different types of quantum phase
  transitions (see Fig.  \ref{fig2} as an example) and reveal genuine quantum
  properties (without classical counterpart) of raw data sets.}
\label{fig1} % Fig 1
\end{figure*}

Here, we show how the full data structure of a path integral (or its related representation as a stochastic series expansion) of certain quantum statistical mechanics models is able to capture genuine quantum effects such as quantum critical behavior~\cite{sachdev_2000} as well as properties of quantum phases. We introduce a stochastic characterization of PI manifolds, and study it in the context of quantum spin models by exploiting state-of-the-art techniques from the field of data mining~\cite{Laio2017}, combined with quantum Monte Carlo (QMC) sampling~\cite{Sandvik2010.2,KrauthBook}. Our results show how very general properties of the path integral manifold - in particular, its intrinsic dimension~\cite{Rozza2015,Laio2017} - display key signatures of quantum critical behavior in several paradigmatic cases, including second-order and Berezinskii-Kosterlitz-Thouless (BKT) quantum phase transitions. This reveals how universal properties do not only dictate simple path integrals properties such as low-order correlations, but, in fact, govern the entire data manifold - signalling, above all, that quantum phase transitions of spin models are accompanied by structural transitions of the corresponding stochastic description of the path integral. At critical points, the path integral representation is parametrically less complex than those of ordered and disordered phases: quantum criticality is thus accompanied by an emergent simplicity in data space, a fact that offers an alternative angle on the representative power of recently developed neural network states~\cite{Carleo2017,Carleo2019,Carrasquilla2020}.

Before continuing, it is worth stressing that, beyond the basic theoretical
goal of characterizing the geometry of  path integrals, our approach is
directly motivated by recent experimental developments in the field of quantum
computing and quantum simulation~\cite{Cirac2012,Nori2020}. While the full
characterization of a many-body wave-function via state tomography is
experimentally prohibitive (when applicable at all), over the last few years
stochastic sampling of wave-functions has become possible in both atomic and
solid state platforms~\cite{Brydges2019probing, Barredo, chiaro2020direct,
Pagano_2020, scholl2020programmable, ebadi2020quantum, Zeiher_2017, Pfau2021}.
In particular, the combination of high-fidelity {\it in situ} imaging
techniques and very fast repetition rates has enabled experiments to collect
thousands of wave-function snapshots, that, as we argue below, are intimately
related to specific types of path integral quantum data sets. These impressive
experimental capabilities have already been exploited in a variety of ways,
including  measurement of entanglement properties and tomography of small
partitions~\cite{Brydges2019probing,chiaro2020direct,elben2020mixed}. Our
approach here differs from these previous attempts, in the fact that we are not
targeting specific wave-function properties (such as entropies), but rather, we
focus solely on extracting universal information by analyzing the data as a
manifold. 

The first element in our analysis concerns the definition of proper 'quantum'
data sets describing the path integral manifold. In a previous
work~\cite{Mendes2020}, some of us have shown how classical partition functions
exhibit very specific patterns in data space, that are universal in the
vicinity of criticality, and where the corresponding structural transition can
be understood utilizing simple arguments based on correlation functions (we
note that, while this manuscript was in preparation, two works have
appeared~\cite{turkeshi2021measurementinduced,schmitt2021observations} that
successfully apply the diagnostic we introduced in Ref.~\cite{Mendes2020} to
specific forms of wave-function representations). Since PIs can be construed as
a highly anisotropic classical partition functions~\cite{Montvay1994}, one
could na\"ively apply methods already developed to understand the latter on the
former.  However, this turns out to be not only a very inefficient formulation
of data mining - as one would have to investigate structural transitions in a
data space where one dimension, the imaginary time, is typically very large -
but, most relevantly, one that will necessarily mix in an uncontrolled manner
real-space and imaginary-time correlations. In addition, the blind application
of the 'classical' procedure will conceal a key aspect of the path integral
representation - namely, the fact that quantum mechanical variables do not
commute. This last aspect will be particularly important in the presence of
non-Abelian symmetries, as we argue below comparing in detail the classical and
quantum cases. One thus needs to identify the proper information to be mined,
and cannot simply translate from the classical case - irrespectively on how
insightful that might be on its own.

We thus introduce two classes of quantum data sets corresponding to the PI
stochastic descriptions [see Fig.~\ref{fig1}(a)]. The first class is obtained
by taking snapshots that are instantaneous in imaginary time, while the second
one incorporates within the data set a finite, discrete fraction of
imaginary-time "slices". We discuss in detail how both of these choices differ
drastically from the equivalent classical description for the two reasons
above. Importantly, both data sets are immediately available when sampling
partition functions via quantum Monte Carlo methods~\cite{Sandvik1997,
becca_sorella_2017, Montvay1994}, and the first one is additionally readily
obtained from experiments, and from the output of exact and variational
wave-function-based methods. 

The second part concerns instead the identification and application of the
proper tools to characterize the complex data structures correspondent to PIs.
The latter are defined in high-dimensional manifolds and may display
non-trivial curvature and topology, in addition to inhomogeneous density
distributions. The first property of the PI manifold we investigate is the
intrinsic dimension - that is, the minimum number of dimensions needed to
accurately describe the manifold itself [see Fig.~\ref{fig1}(b)]. This allows
us to minimally characterize the sampled PI manifold with a single number, that
can be efficiently estimated with state-of-the-art algorithms.

For all the models considered here, the intrinsic dimension displays a minimum
at transition points - indicating that the geometry of the PI manifold {\it
simplifies} at criticality. This observation points to the fact that the PI
manifold behaves independently from quantum correlations such as bipartite
entanglement, that are often maximal at quantum critical points
(QCPs)~\cite{Calabrese_2009, Amico2008}; our findings are instead suggestive of
the fact that the PI data structure is inheriting simplicity from the fact that
the low-energy properties are captured by very few degrees of freedom (DOF) -
and thus, constrain the PI structure. This is reminiscent of the fact that
energy spectra are also highly constrained by universal properties: however,
while the latter typically depend solely on low-order correlation functions,
the constrained structure we observe is related to arbitrary order
correlations.

\begin{figure}[t]
\begin{center}
{\centering\resizebox*{8.8cm}{!}{\includegraphics*{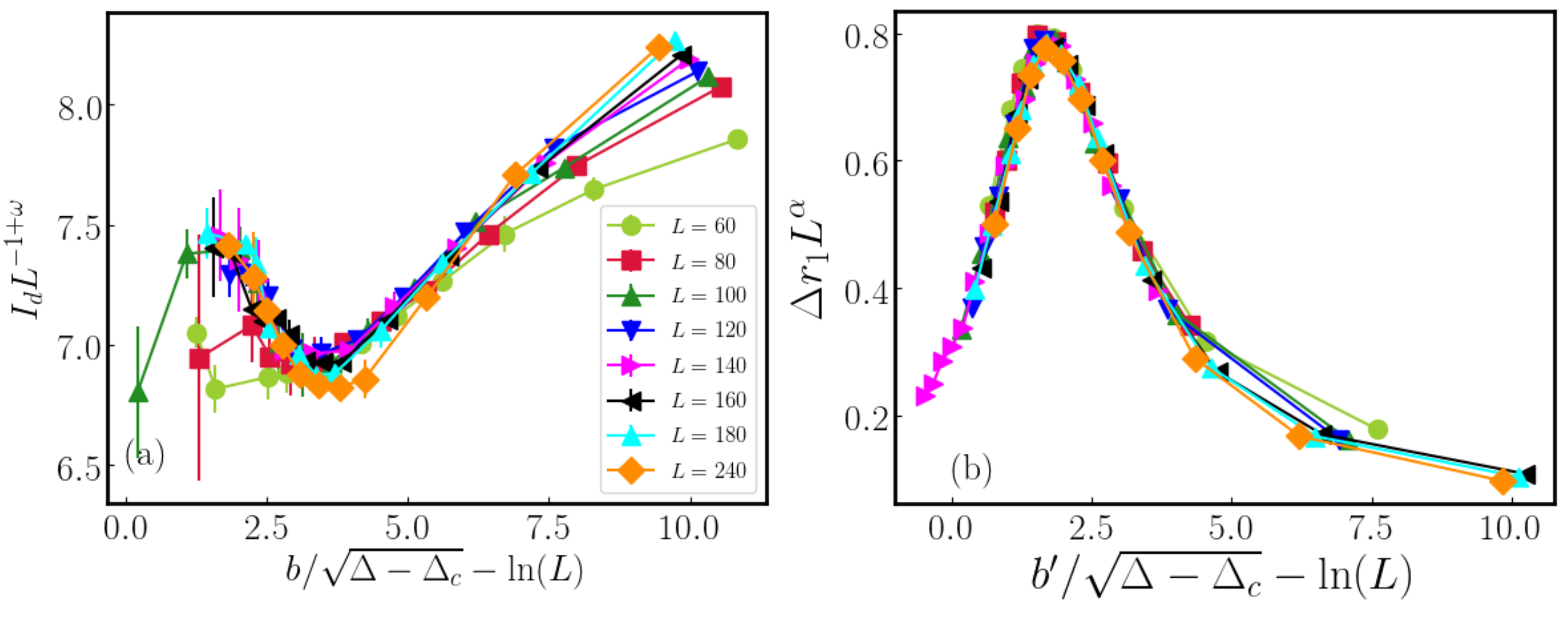}}}
\end{center}
\caption{As a representative example of our results, we show the universal data
  collapse of both the $I_d$ [panel (a)] and $\Delta r_1$ [panel (b)] for the
  one-dimensional $1/2$-XXZ model in the vicinity of the
  Berezinskii-Kosterlitz-Thouless transition (see text).}
\label{fig2} % Fig 2
\end{figure}

The relation between PI data structure and universal behavior becomes apparent
when performing a finite-size scaling (FSS) analysis of the intrinsic dimension
\cite{Sandvik2010.2}: the latter displays universal scaling collapse, whose
functional form is dictated by the universality class (second-order or BKT),
and by its critical exponent $\nu$. An example of such scaling collapses is
illustrated in Fig.~\ref{fig2}(a) for the case of BKT transition in the
one-dimensional (1D) XXZ spin model, which we investigate below.

In addition to the intrinsic dimension, we also investigate the statistical
properties of the distances between nearest-neighbor (NN) configurations in the
path integral data space. At the qualitative level, the corresponding
distribution describes fluctuations of configurations within the path integral
manifold: the broader it is, the stronger we expect correlations in our data
set - that is, correlations in both space and imaginary time. We show how this
quantity - of simpler experimental and computational access than the intrinsic
dimension - also provides key signatures of critical behavior [see
Fig.~\ref{fig2}(b)]. Moreover, beyond the critical regime, we show that it
reveals fundamental properties of symmetry-broken phases.

On general grounds, our work aims to learn quantum phases and their transitions in a purely unsupervised manner, i.e., without prior knowledge of the system (e.g., the nature of phases or the phase transitions).
While a series of works have recently adopted similar approaches
~\cite{LeiWang2016,Scalettar2017,Scheurer2019,Singh2017,Eshan2018,Chaves2019,Yang2020,Lidiak2020,Nori2020,Yoshihiko2020,Acin2020,Wei2021}, it is worth mentioning that our route is complementary, and to some extent alternative, to these works, as we access data-based quantities without any projection or compression of the data set.
Our approach consists of data mining data sets as a whole, which, as we will illustrate below, allows to access universal properties of quantum systems (while other properties might not be retrievable).

The paper is structured as follows. In Sec.~\ref{quantumdataset}, we review the
formulation of quantum partition functions using path integrals and stochastic
series expansions: this allows us to fix notations, and to define the quantum
data sets we are interested in. In Sec.~\ref{sec:qdata}, we introduce the tools
we adopt to analyze quantum data sets (i.e., the intrinsic dimension and the
statistics of distances) and review the estimators we employ. In
Sec.~\ref{sec:result1} and Sec.~\ref{sec:result2}, we discuss our analysis of
second-order and BKT quantum critical points, respectively. In
Sec.~\ref{sec:sym}, we investigate the role of non-Abelian symmetries, and
contrast it to the classical case. Finally, in Sec.~\ref{sec:concl}, we draw
our conclusions.

\section{Quantum data sets and models}
\label{quantumdataset}

In order to investigate the properties of our quantum many-body systems of
interest, we perform quantum-to-classical mappings which transform the original
configurational space of the quantum model into an equivalent classical one,
amenable to simulations via appropriate Monte Carlo (MC) algorithms.
Subsequently, we analyze the data set of the configurations sampled during the
MC simulations via the data-mining-inspired observables mentioned above.

In this section, we offer a discussion of the Hamiltonians studied in our work,
with particular focus on their ground-state (quantum) critical behavior,
followed by an introduction to the employed quantum-to-classical mappings.

\subsection{Models}
\label{IIa}

We analyze the Transverse Field Ising Model (TFIM; see, e.g., \cite{sachdev_2000} for a thorough introduction)
\begin{equation}    \label{eq.ham-tfim}
  H = -2h \sum_{i = 1}^L S_i^x - 4 \sum_{\langle i,j \rangle} S_i^z S_j^z
\end{equation}
for a one-dimensional system of size $L$ with periodic boundary conditions
(PBC), where $S_i^x$ ($S_i^z$) denotes the $x$ ($z$) component of the quantum
spin-$1/2$ operator $\boldsymbol{S}_i$ acting on site $i$, and $\langle i,j
\rangle$ denotes a sum over NN pairs of sites. In the ground-state regime, the
TFIM undergoes a second-order phase transition at the critical field value $h_c
= 1$ between a paramagnetic and a ferromagnetic (FM) state, corresponding to $h
> h_c$ and $h < h_c$, respectively. Sec. IV A focuses on the determination of
the position of the TFIM critical point via data mining of the configurational
data set.

We also study the spin-$1/2$ Heisenberg Bilayer model \cite{Sandvik1994,Sandvik2006}, described by the
Hamiltonian
\begin{equation}    \label{eq.ham-bilayer}
  H = \sum_{a = 1,2} \sum_{\langle i,j \rangle} \boldsymbol{S}_{i,a} \cdot
  \boldsymbol{S}_{j,a} + g \sum_{i = 1}^{N_s} \boldsymbol{S}_{i,1} \cdot
  \boldsymbol{S}_{i,2},
\end{equation}
where the layer index $a$ identifies one of two symmetrical square lattices of
$N_s = L \times L$ sites composing the bilayer geometry, and $g$ is the
relative strength of the interlayer onsite Heisenberg term with respect to the
intralayer NN one. In the ground-state regime, the model displays an
$SU(2)$-symmetry-broken antiferromagnetic (AFM) phase and an $SU(2)$-disordered
phase for weak and strong $g$, respectively. The transition between these
states has been shown to belong to the three-dimensional $O(3)$ (Heisenberg)
universality class, and to be characterized by a value of the critical
parameter $g = g_c = 2.5220(1)$, and a correlation length critical exponent
$\nu = 0.7106(9)$ \cite{Sandvik2006}. The determination of these two quantities
via unsupervised learning of the QMC configurational data sets is discussed in
Sec. IV B.

Finally, we investigate the one-dimensional spin-$1/2$ XXZ Hamiltonian \cite{Gogolin2004, Giamarchi2004}
\begin{equation}    \label{eq.ham-xxz}
  H = \sum_{i = 1}^{N_s} \left( S_i^x S_{i + 1}^x + S_i^y S_{i + 1}^y + \Delta
  S_i^z S_{i+1}^z \right).
\end{equation}
In the $\Delta > 0$ region, the phase diagram of the model displays a gapless
critical phase for $\Delta < 1$, separated from a
$\mathbb{Z}_2$-symmetry-broken antiferromagnetic phase by a
BKT critical point at $\Delta = 1$~\cite{Gogolin2004}. The
features of this kind of phase transition are radically different from those of
second-order criticality such as the examples mentioned above, and include,
e.g., the non-locality of the order parameter associated to the transition, as
well as an exponential (rather than power-law) divergence of the correlation
length in proximity of the transition. The results of our application of
data-mining-related observables to the study of this kind of critical behavior
in the one-dimensional XXZ model are discussed in Sec. V.

\subsection{Quantum-to-classical mappings and quantum data sets}
\label{IIb}

The partition function of a quantum system described by a Hamiltonian $H$ is
\begin{equation}    \label{eq.partfunc}
  Z = \sum_{\left \{ \ket{\alpha} \right \}} \bra{\alpha} e^{-\beta H}
  \ket{\alpha},
\end{equation}
where $\beta = 1/T$ is the inverse temperature of the system (in units of the
Boltzmann constant), and $\left \{ \ket{\alpha} \right \}$ is a complete basis
set for the Hilbert space in which $H$ operates.
In the following, we will employ a basis set in terms of the eigenvalues of $S^z$ operators: $\ket{\alpha} = (S_{1}^z,...,S_{N_s}^z)$ for all models considered.
In the MC approaches considered in this work, many-body configurations are
sampled with a probability proportional to their contribution to the partition
function, making the evaluation of terms such as those in the sum in Eq.~\eqref{eq.partfunc} crucial.

The first of the MC techniques adopted in this work is known as \textit{Path Integral Monte Carlo} (see, e.g., \cite{Sandvik1997}), and allows a direct, unbiased sampling of the Path Integral manifold. In this approach, the Hamiltonian of the original quantum system is
decomposed as $H \equiv H_0 + H_1$, where the two terms are diagonal and
non-diagonal on the basis set formed by the $\ket{\alpha}$, respectively: the
partition function in Eq. \eqref{eq.partfunc} can then be rewritten as
\begin{align}
  & Z = \sum_{\left \{ \ket{\alpha} \right \}} \bra{\alpha} \prod_{k = 0}^{M -
  1} e^{- \epsilon \left( H_0 + H_1 \right)} \ket{\alpha}
  \label{eq.pf-pimc-partial} \\
  & = \sum_{\left \{ \ket{\alpha_{k = 1, \ldots, M}} \right \}} \prod_{k = 0}^{M - 1} \bra{\alpha_k} e^{-\epsilon H_0} e^{-\epsilon H_1} \ket{\alpha_{k + 1}} + O
  \left( \epsilon^2 \right) \nonumber \\
  & = \sum_{\left \{ \ket{\alpha_{k = 1, \ldots, M}} \right \}} \prod_{k = 0}^{M - 1} e^{-\epsilon \bra{\alpha_k} H_0 \ket{\alpha_k}} \bra{\alpha_k} e^{-\epsilon
  H_1} \ket{\alpha_{k + 1}} + O \left( \epsilon^2 \right),
  \label{eq.pf-pimc}
\end{align}
where $\epsilon \equiv \beta/M$, each of the sets $\left \{ \ket{\alpha_k} \right \}$ is a complete set for the Hilbert space, and $\ket{\alpha_M} \equiv \ket{\alpha_0}$.

For quantum spin systems, such as those considered in our work, the states $\ket{\alpha_k}$, also known as \textit{slices}, are usually chosen as eigenvectors of the $z$-component of the quantum spin-$1/2$ operators $\boldsymbol{S}_i$ acting on each site $i$, i.e., $\ket{\alpha_k} \equiv \ket{S_{1k}^z, \ldots, S_{N_s k}^z}$, where $N_s$ is the system size. The union of the various sets $\left \{ \ket{\alpha_k} \right \}$ in Eq. \eqref{eq.pf-pimc} can be interpreted as an
extended configuration space, whose dimensionality is increased by 1 with
respect to the one of the original quantum system, and whose MC sampling can be performed via a conventional Metropolis algorithm due
to the relative simplicity of the calculation of the matrix elements in Eq. \eqref{eq.pf-pimc}. In the following, we adopt the PIMC technique to analyze
the one-dimensional TFIM described by the Hamiltonian in Eq. \eqref{eq.ham-tfim}, which is mapped by the procedure discussed above to a two-dimensional, classical, anisotropic Ising model. The configuration space of
the latter is then sampled via the use of conventional Wolff cluster updates
\cite{Wolff1990}.

As discussed in the Introduction, our approach is based on the analysis of
generic features of the data structure associated with classical
representations of quantum partition functions [the definition of the data
structure is shown in Fig. \ref{fig1}(a)]. The PI represents one choice for
such representation.  However, to demonstrate the flexibility of our procedure,
we analyze some of our models of interest using an equivalent and related
representation to the PI one: namely, the \textit{Stochastic Series Expansion}
(SSE) approach (see, e.g.,~\cite{Sandvik1991}).

In this method, the quantum partition
function in Eq. \eqref{eq.partfunc} is rewritten by expanding the exponential
operator in power series of $H$
\begin{align}
  & Z = \sum_{k = 0}^{\infty} \frac{(-\beta)^k}{k!} \sum_{\left \{ \ket{\alpha}
  \right \}} \bra{\alpha} H^k \ket{\alpha} \nonumber \\
  & = \sum_{k = 0}^{\infty} \frac{(-\beta)^k}{k!} \sum_{\left \{ \ket{\alpha_k}
  \right \}} \bra{\alpha_0} H \ket{\alpha_{k-1}} \times \bra{\alpha_{k-1}}
  H \ket{\alpha_{k-2}} \times ... \nonumber \\
  & \times \bra{\alpha_2} H \ket{\alpha_1} \times \bra{\alpha_1} H
  \ket{\alpha_0},      \label{eq.pf-sse}
\end{align}
where the definition and constraints on the $\left \{ \ket{\alpha_k} \right \}$ are the same as in Eq.~\eqref{eq.pf-pimc}. As in the PI case, the ensemble of the $\left \{ \ket{\alpha_k} \right \}$ can be interpreted as a higher-dimensional configuration space; a schematic representation of the latter as obtained by applying the SSE (or PI) mapping is displayed in Fig.~\ref{fig1}(a).

As all the matrix elements in Eq.~\eqref{eq.pf-sse} are easily evaluated (and positive-definite in the case of the models considered here), the configuration space is now amenable to importance MC sampling according to the partition function, which in the following is performed via a combination of diagonal and off-diagonal
directed-loop updates \cite{Sandvik1999}.

The two mappings described above can be proven to be identical in the limit $M
\to \infty$ \cite{Sandvik2019-1}. On the one hand, if one takes the SSE
partition function in Eq. \eqref{eq.pf-sse}, chooses a ``cut-off'' $k_{\max} =
M$ for the expansion order, and adds $(M - k)$ matrix elements of the identity
to each term of order $k$, one obtains
\begin{align}
  & Z_{\mathrm{SSE}} = \sum_{\left \{ S_M \right \}}  \sum_{\left \{
    \ket{\alpha_k} \right \}} \frac{\beta^k (M - k)!}{M!} \bra{\alpha_0}
  H_{i_M} \ket{\alpha_{M - 1}} \times \ldots   \nonumber \\
  & \times \bra{\alpha_2} H_{i_2} \ket{\alpha_1} \bra{\alpha_1} H_{i_1}
  \ket{\alpha_1},    \label{eq.pf-sse-equiv}
\end{align}
where $\left \{ S_M \right \}$ identifies the ensemble of all sequences of
length $M$ of operators $H_i$, the indices $i = 0,1$ denotes the identity or
the Hamiltonian operator, respectively, and the combinatorial factor $M!/
\left[ k!  (M - k)!  \right]$ has been introduced to keep into account the
equivalent ways to insert the $(M - k)$ identity operators in the $M$-sized
operator string.

On the other hand, the partition function in Eq. \eqref{eq.partfunc} can be
rewritten, starting from Eq. \eqref{eq.pf-pimc-partial}, as
\begin{align}
  & Z_{\mathrm{PI}} = \sum_{\left \{ \ket{\alpha_k} \right \}} \bra{\alpha_0} 1 -
  \epsilon H \ket{\alpha_{M - 1}} \times \ldots     \nonumber \\
  & \times \ldots \times \bra{\alpha_2} 1 - \epsilon H \ket{\alpha_1}
  \bra{\alpha_1} 1 - \epsilon H \ket{\alpha_0} + O \left( \epsilon \right)
\end{align}
The terms can then be rearranged, using the notations introduced in Eq.
\eqref{eq.pf-sse-equiv}, as
\begin{align}
  & Z_{\mathrm{PI}} = \sum_{\left \{ S_M \right \}} \sum_{\left \{ \ket{\alpha_k}
  \right \}} \epsilon^k \bra{\alpha_0} H_{i_M} \ket{\alpha_{M - 1}} \times
  \ldots     \nonumber \\
  & \times \ldots \times \bra{\alpha_2} H_{i_2} \ket{\alpha_1} \bra{\alpha_1}
  H_{i_1} \ket{\alpha_0} + O \left( \epsilon \right)    \label{eq.pf-pimc-equiv}
\end{align}
In the limit $M \to \infty$, the prefactor $\beta^k (M - k)!/M!$ in
\eqref{eq.pf-sse-equiv} converges to $\beta^k/M^k = \epsilon^k$, while the
approximation error in \eqref{eq.pf-pimc-equiv} vanishes, implying the
equivalence of the expressions for $Z$ obtained in the two formalisms. In our
calculations, with both the PI and SSE approach, we increase the number of
slices until finite-$M$ effects are negligible, ensuring the attainment of the
limit mentioned above, and therefore the equivalence of the PI and SSE methods
to analyze our problems of interest in the regimes investigated in this study.

The data sets analyzed in our work are composed by stochastically sampled
elements of the extended configuration spaces discussed above, written in terms of the the set of slices $\{\ket{\alpha_i}\}$. More specifically, we consider sets of either 
\begin{itemize}
\item single-slice configurations $\vec{X} = \left \{ \ket{\alpha_0} \right \}$, or
\item  configurations containing a subset of
$M' \leq M$ evenly spaced slices, i.e., $\vec{X}  = \left \{ \ket{\alpha_{k_1}},...,\ket{\alpha_{k_{M'}}}  \right \}$, where $k_i \equiv i \times \left[ M/M' \right]$, $i = 0, \ldots, M' - 1$, and $[x]$ denotes the integral part of a real number $x$ (see Fig.~\ref{fig1}).
\end{itemize}
As anticipated in the introduction, the first data set also corresponds to
wave-function snapshots in experiments with \textit{in situ} imaging, while the
second data set genuinely displays path integral features, incorporating
effects from imaginary-time correlations.

\section{Data mining quantum data sets}\label{sec:qdata}

We investigate generic features of data sets aiming to extract useful
information about quantum criticality. More specifically, we consider basic
data set features associated with the statistics of distances between
neighboring configurations: namely, the intrinsic dimension and the variance of
the distribution of neighbouring distances. Below, we describe in more detail
the key quantities and discuss their connections with physical properties of
quantum phase transitions.

\subsection{Intrinsic dimension and two-NN estimators}  
\label{id_nn}

A common way to deal with data sets is to consider each data instance as a point in a space whose dimension (the embedding dimension, $N_c$) is the number of features needed to describe each sample. However, the existence of correlations between data points usually leads to situations in which the points live, approximately, in a manifold whose dimension, known as \textit{intrinsic dimension} (ID) and denoted by $I_d$, is much lower than $N_c$. The basic intuition behind the $I_d$ is illustrated in Fig. \ref{fig1}:
although the synthetic  data set of panel (b) is embedded in a 3D space, its essential content can be described (\textit{almost} without loss of information) by a non-linear manifold whose $I_d$ is equal to 1. In simple cases like this, the $I_d$ corresponds to the
minimum number of variables needed to describe a data set.

Different approaches have been proposed to estimate the $I_d$; see Ref.
\cite{Rozza2015} for an extensive discussion about this topic. The technique
used here, the TWO-NN~\cite{Laio2017}, is based on a class of methods that
relies on the statistics of distances between NN elements in the data set. The
basic idea of such approaches is that nearest-neighborhood points can be
considered as uniformly drawn $I_d$-dimensional hyperspheres
\cite{Rozza2015,Bickel2005}. This assumption allows one to establish relations
between the $I_d$ and the statistics of neighboring distances. In particular,
in the TWO-NN, for each point $\vec{X}$ in the data set [see Fig. \ref{fig1}
(b)] one considers its distance from its NN and next-nearest-neighbor (NNN)
point $r_1(\vec{X})$ and $r_2(\vec{X})$, respectively. The set of distances
$r_1$ and $r_2$ are defined in terms of the Euclidean distance (see below).
Under the condition that the data set is locally uniform in the range of
next-nearest-neighbors, it has been shown in Ref.~\cite{Laio2017} that the
formula for the distribution function of ${\mu = r_2(\vec{X})/r_1(\vec{X})}$ is
\begin{equation}
 f(\mu) = I_d \mu^{-I_d - 1}
 \label{eq.pdfmu}
\end{equation}
or, in terms of the cumulative distribution $P(\mu)$,
\begin{align}
 I_d = - \frac{\ln\left[ 1 -  P(\mu) \right]}{\ln\left( \mu \right)}.
 \label{Id}
\end{align}
Due to the minimal extension of the neighborhood considered, the TWO-NN method
is particularly suitable for non-linear manifolds, which is important when
dealing with physical data sets \cite{Mendes2020}.

It is worth mentioning that the TWO-NN is designed for configurations defined
on a continuous support. However, the generalization to configurations
describing discrete data sets, such as those considered in this work, is
straightforward and does not display problems if a large enough number of
coordinates $N_c$ is considered.

Before proceeding, let us define more precisely the quantum data sets
introduced at the end of last section. As described in Fig. \ref{fig1}, such
data sets are defined by a set of points $\vec{X}^i = (X^i_1, X^i_2, ...,
X^i_{N_c})$, where the index $i$ is the label  of the configurations sampled in
the MC simulations (i.e., $i = 1, ..., N_r$), $N_c$ is the number of
coordinates (or the embedding dimension, as mentioned above), and $N_r$ is the
total number of points considered in the data sets. The coordinates $X^i_j$ are
defined in terms of the path integral (or stochastic series expansion) degrees
of freedom, as explained in Fig. \ref{fig1}(a) and in Sec. \ref{IIb}.

\subsection{Scale dependence of the $I_d$ and the statistics of neighbouring distances}
\label{corre_id}

One key aspect is that the $I_d$ (computed via the TWO-NN method) is a
scale-dependent quantity. More specifically, the $I_d$ is measured on a range
scale defined by the NN and NNN distances $r_1$ and $r_2$ \cite{Laio2017}. For
each point in the phase diagram, the scale is determined by the total number of
points in configuration space, $N_r$, since the latter fixes the average values
of $r_1$ and $r_2$. Indeed, the effect of changing $N_r$ is analogous to
zooming in or zooming out the data set in configuration space, which changes
the value of $I_d$ [see the pictorial example in Fig. \ref{fig1}(b)]. The
$I_d$ also reveals changes of scale associated with structural transitions in
configuration space.  For example, we mention the structural transitions
occurring in classical data sets in proximity of thermal phase transitions
\cite{Mendes2020}.

The fundamental reason why the $I_d$ exhibits a singular behavior in the
vicinity of classical transitions is related to changes of scale in
configuration space \cite{Mendes2020}, which interestingly can be associated to
significant changes in the physical properties of the connectivity of
neighboring points. For example, in the case of the classical Ising transition
the data structure related to Ising ferromagnetic phases is characterized by
configurations whose NN and NNN have essentially equivalent physical properties
(e.g., magnetization): conversely, in the disordered phase the physical
properties of neighboring configurations are entirely uncorrelated. A similar
reasoning applies in the case of a the classical BKT transition, where
neighboring points are characterized by configurations with the same
topological properties (i.e., winding number) in the ordered phase.

The reasoning mentioned above serves as a guideline for defining other
quantities associated with the statistics of neighboring distances.  As we
discuss below, such quantities (going beyond the $I_d$) reveal essential
properties of path integral data sets. More specifically, we consider the
distribution function associated to the NN (NNN) distances, and its variance
$\Delta r_1$ ($\Delta r_2$):
\begin{equation}
  \Delta r_i = \left< r_i^2 \right> -  \left< r_i \right>^2,
\end{equation}
where $\left< r_i \right> = N_r^{-1} \sum_{k=1}^{N_r} r_i(\vec{X^k})$, (with $i =
1,2$); $r_1(\vec{X^k})$ and $r_2(\vec{X^k})$ are the first and the second nearest-neighbor distances associated to the configuration $\vec{X^k}$, respectively. At least in the case where the data sets are homogeneous in density, the
$\Delta r_i$ can detect changes of scale in configuration space, similarly to
the $I_d$. Furthermore, the $\Delta r_i$ reflects the global connectivity of
neighboring points in configuration space, which is a fundamental ingredient to
detect topological transitions \cite{Scheurer2019}. We note that the $\Delta
r_i$, differently from $I_d$, are also sensitive to inhomogeneity in the
sampling of the data set: therefore, in general we expect $I_d$ to provide a
more reliable description of universal properties.

\subsection{Distances and correlations}

A crucial step to obtain both the $I_d$ and $\Delta r_i$ is to consider a
proper metric. Here, we compute the distance $r(\vec{X}^i, \vec{X}^j)$ between
two configurations $\vec{X}^i$ and $\vec{X}^j$ according to the well-known
Euclidean metric, following which the distance can be straightforwardly recast
in the form
\begin{equation}
  r(\vec{X}^i, \vec{X}^j) = \sqrt{2 N_c \left( 1 - \frac{1}{N_c}
  \sum_{p=1}^{N_c} X^i_p X^j_p \right)}.
 \label{disE}
\end{equation}
This choice satisfies the basic requirements for a proper metric: namely, it is
non-negative, equal to zero only for identical configurations, symmetric, and
it respects the triangular inequality.

It is worth mentioning that the Hamming distance is also a proper metric, which could be used in place of the Euclidean one leading to only a trivial change in the intrinsic dimension for the data sets
considered in our work (i.e., binary variables): more specifically, $I_d^E = 2 I_d^H$, where $I_d^E$ and $I_d^H$ are the intrinsic dimension computed with the Euclidean and the Hamming metric, respectively. This result can be understood if we (i) define the Euclidean and Hamming distances between two configurations as $r^E = 2\sqrt{N_{\mathrm{diff}}}$ and $r^H = N_{\mathrm{diff}}$,
where $N_{\mathrm{diff}}$ is the number of unequal coordinates of the configurations, and (ii) consider that the ID is a function of $\ln(r_2/r_1)$ (see Eq. \eqref{Id}).

An advantage of considering a standard metric (such as the Euclidean distance)
is that we can efficiently compute the set of NN distances $r_1$ and $r_2$ with
state-of-the-art unsupervised NN search algorithms, leading to $O \left( N
\log(N) \right)$ scaling of the computational complexity \cite{scikit-learn}.
However, it is worth mentioning that $r(\vec{X}^i,\vec{X}^j)$ does not reflect
the underlying symmetries of the physical configurations $\vec{X}$. For
instance, here we consider systems with translation symmetry (in space and
imaginary-time direction), for which the distance between two configurations
$\vec{X^i}$ and $\vec{X^j}$ related by a given translation symmetry operation
should be equal to zero. As a simple example, let us consider $\vec{X^i} =
(-1,-1,-1,1)$ and $\vec{X^j} = (-1,1,-1,-1)$: despite these configurations being
physically equivalent by translational symmetry, we have $r(\vec{X}^i,
\vec{X}^j) > 0$. Nevertheless, this drawback does not affect our results, given
that the probability of sampling two or more configurations belonging to the
same translational symmetry sector is exponentially suppressed for large number
of configurational DOF (typically, we consider $N_c > 100$).

One key aspect of the form of $r(\vec{X}^i, \vec{X}^j)$ displayed in Eq.
\eqref{disE} is that it reveals the intimate relation between generic data set
features and correlations described by the terms $X^i_p X^j_p$. An interesting
perspective is to try to connect such data correlations with correlations
between the variables themselves, i.e., correlations of the type $X^i_p X^i_r$,
where $p$ and $r$ are indices related to the coordinates of a given
configuration vector $i$. From now on, we call the later \textit{physical
correlations}; for the quantum data sets considered here, the indices $p$ and
$r$ may be separated by distances in both space and/or imaginary time.
Although it is hard to establish this connection in general, one can verify
that it indeed holds in certain temperature regimes of classical systems
\cite{Mendes2020}, which explains (at least qualitatively) why the $I_d$ and
quantities related to the statistics of neighboring distances exhibit universal
scaling behavior in the vicinity of classical critical points. However, whether
or not such a connection holds for quantum systems, or if generic features of
the data sets are related to universal properties of quantum critical points,
cannot be immediately answered based on the classical results. As we discuss
now, quantum data sets differ in a fundamental way from their classical
counterparts.

\subsection{Differences between quantum and classical partition-function data sets}
\label{classical_quantum}

The path integral of a $D$-dimensional partition function can be mapped to a (highly-anisotropic) classical partition function in $D+1$ dimensions. It is thus natural to wonder whether one could just employ the same methods already applied to study the latter in the context of the former. 
The answer is negative for three main reasons - two of conceptual and one of practical nature. 

The first limitation is that analyzing PIs as classical data sets will
necessarily mix information contained in space and imaginary-time correlations:
this will make it hard to identify precise connections between the data
structure itself and physical phenomena, as it will correspond to analyzing
arbitrary space- and time- correlations. The reason why identifying such
connections would be challenging is that physical information (such as critical
exponents) is typically referred to specific correlation functions in either in
space or in time: for instance, the correlation length critical exponent $\nu$
is associated to equal-time correlators, while certain anomalous critical
exponents are related to the decay of single-site Green functions in imaginary
time. This is in sharp contrast with the classical case, where correlations are
isotropic. Analyzing the full PI manifold would unavoidably mix these two types
of relevant information.

The second limitation is that only a given part of the PI manifold is
experimentally accessible. This is, to the best of our knowledge, a consequence
of the fundamentally quantum nature of the problem: once wave-function
snapshots are taken, these are necessarily in the form of strong measurements.

The third limitation is of practical nature. Data mining a manifold becomes
impractical when the dimension of the embedding data space increases. {\it A
priori}, this does not seem an issue, as the number of points necessary to
characterize the intrinsic dimension of a manifold is typically related to the
dimension of the manifold itself, and not to that of the embedding space. A
simple example of this fact is the identification of a line in a
$D$-dimensional space, for which one just needs a number of points that scales
with the intrinsic dimension. However, in practical terms, one still requires a
larger number of samples to properly characterize the manifold features,
especially for the very large values of the intrinsic dimensions we will
encounter below.

The three considerations above highlight the fact that one cannot simply take
the same approach demonstrated with classical partition functions, and apply it
to PI manifolds: in the best case, this would lead to a hard-to-decipher and
experimentally inaccessible picture, while in the worst-case scenario the
classically-mutuated approach will be simply inapplicable. The quantum data
sets described above overcome these limitations at different levels, via either
focusing on a single slice, or capturing imaginary-time properties in a
selective manner. 

There is an additional, genuinely quantum-mechanical aspect that quantum data
sets have to handle: namely, the fact that quantum fields do not commute. Since
the definition of the data set requires specifying a given basis, this will
inevitably lead to new features in the presence of non-Abelian symmetries, as
the latter are characterized by non-commuting conserved quantities. In
Sec.~\ref{sec:sym}, we will address this specific aspect in the context of
SU(2) symmetries. 

\section{Second-order transitions}
\label{sec:result1}

We begin by considering the behavior of both the $I_d$ and the NN distance
distribution variance $\Delta r_1$ in the vicinity of two paradigmatic examples
of second-order quantum phase transitions.

\subsection{One-dimensional quantum Ising model}
\label{IVa}

In this section we discuss the results of the application of the observables
introduced in Sec. III to the analysis of the quantum critical behavior of the
one-dimensional TFIM (see Sec. IIB). The required configuration data sets are
generated via PIMC simulations performed at inverse temperature $\beta = 512$,
where convergence in temperature to the ground state regime was observed for
the order parameter associated to the ferromagnetic transition [i.e., the
squared magnetization $m_z^2 = \frac{4}{L^2} \left( \sum_{i = 1}^{L} S_i^z
\right)^2$]. All the simulations considered in the following are performed with
$M = 131072$ slices, of which $M' = 512$ are considered in the configurations
which compose the analyzed data set.

In order to ensure a thorough enough sampling of the $L \times M'$ DOF in each of the configurations, we sample $N_r = 32768$ configurations for each of our simulations, which is more than or equal to twice the number of DOF in any of the cases we analyze. These configurations are sampled at regular intervals in MC time (i.e., the distance between them in number of updates is
constant). To avoid correlated sampling, for each of our simulations we compute the autocorrelation time of one of the DOF (assuming a weak dependence of such a quantity on this choice) and continue the simulation until it is possible to extract $N_r$ configurations such that the distance in simulation time between
them is larger than or equal to the estimated autocorrelation time.

\begin{figure}[t]
  \begin{center}
    {\centering\resizebox*{8.8cm}{!}{\includegraphics*{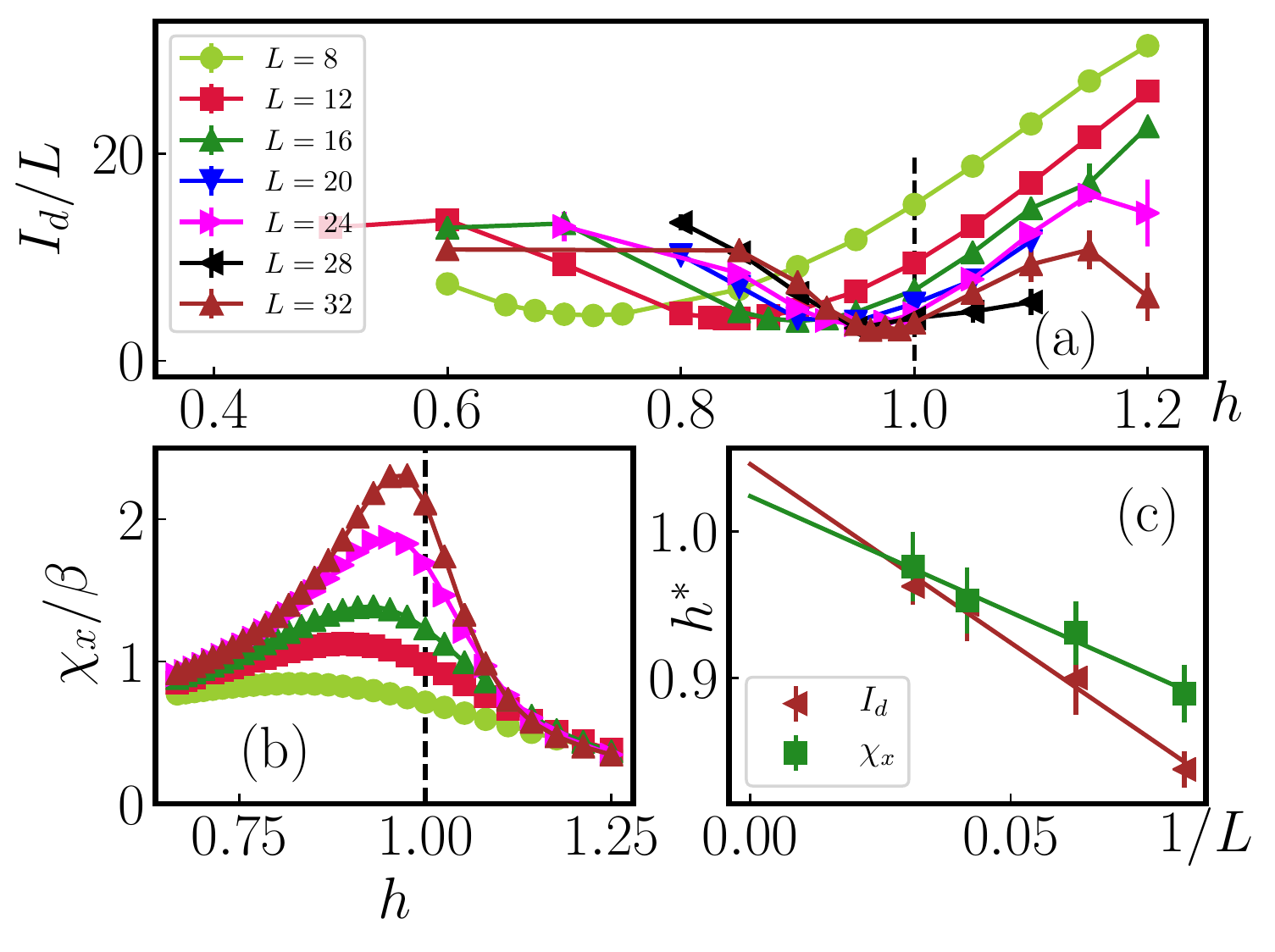}}}
  \end{center}

  \caption{\textit{Quantum Ising model (Intrinsic dimension).} Panel (a):
  intrinsic dimension $I_d$ in units of the system size $L$ as a function of
  $h$. Panel (b): magnetization variance in the $x$ direction as a function of
  the transverse field $h$. Panel (c): extrapolation to the thermodynamic limit
  of the position of the $I_d$ minima $h^*(L)$ and of the susceptibility peaks
  $h_{\chi}^*(L)$ as a function of $1/L$; we obtain $h^\infty = 1.045(7)$ and $h^{\infty}_{\chi} = 1.02(1)$, respectively. In all panels, the vertical dashed
  line corresponds to the critical point $h = h_c$.}

  \label{figIDIsing} % Fig 1
\end{figure}

The first step in our analysis is the direct calculation of the $I_d$ of the
sampled configurations. The results for different system sizes are shown in
Fig.~\ref{figIDIsing}(a) as a function of the transverse field $h$. The most
striking feature of the behavior of the $I_d$ here is the presence of a minimum
at a size-dependent value of the field $h^*(L) < h_c$, which for larger sizes
progressively moves towards the critical point. We perform a linear fit of the
position of this minimum in the thermodynamic limit $L \to \infty$ [see brown
triangles in Fig.~\ref{figIDIsing}(c)] obtaining an extrapolated value
$h^{\infty} = 1.045(7)$.

In order to compare the accuracy of the $I_d$ estimate for the transition point with that obtainable via conventional FSS analysis, we
compute the variance of the magnetization distribution along the $x$ axis
\begin{equation}
  \chi_x \equiv L \beta \left( \ave{m_x^2} - \ave{\left| m_x \right|}^2 \right),
\end{equation}
where $m_x \equiv \frac{2}{L} \sum_{i = 1}^L S_i^x$. This quantity can be
calculated by exploiting the self-duality of the one-dimensional TFIM under an
axis rotation mapping the $x$ axis into the $z$ axis. In particular, this
property results in the identity between $m_x$ computed at a value $h$ of the
transverse field and $m_z$ computed at $h' = 1/h$, with the latter being
straightforward to obtain in the PIMC approach.

The observable $\chi_x$ is reminiscent of the susceptibility for a classical
$\mathbb{Z}_2$-symmetry-breaking phase transition. Indeed, it also shares a
similar finite-size behavior (see, e.g., \cite{Sandvik1999}), peaking at a
size-dependent value $h^*_{\chi}(L) < h_c$ of the transverse field which
approaches the exact critical point as the size increases [see Fig.
\ref{figIDIsing}(b)], a behavior which, as discussed above, is also displayed
by the $I_d$. A linear extrapolation of the finite-size peak positions as a
function of the inverse size [see green squares in Fig. \ref{figIDIsing}(c)]
returns an extrapolated value $h^{\infty}_{\chi} = 1.02(1)$. The latter
displays comparable accuracy to the $I_d$ estimate obtained using the same system
sizes, proving the substantial equivalence in precision between FSS of the
$I_d$ and that of more conventional observables usually associated to critical
behavior.

\begin{figure}[t]
  \begin{center}
    {\centering\resizebox*{8.8cm}{!}{\includegraphics*{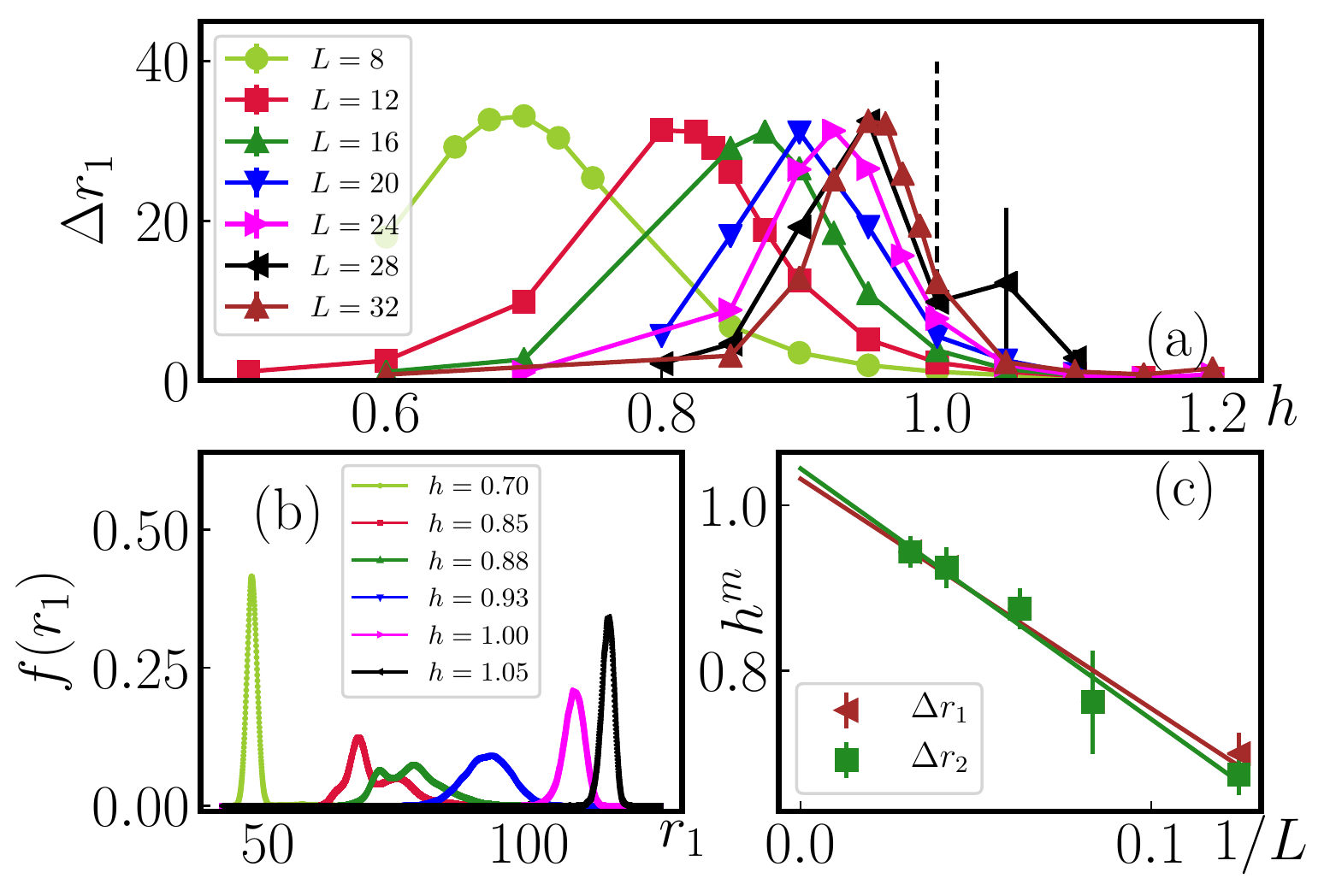}}}
  \end{center}

  \caption{\textit{Quantum Ising model (statistics of NN distances).} Panel
  (a): variance $\Delta r_1$ of the distribution function of the NN distances
  $f(r_1)$ as a function of $h$. Panel (b): $f(r_1)$ for $L = 16$ as a function
  of $h$. Panel (c): extrapolation to the thermodynamic limit of the position
  of the maxima $h^m(L)$ of $\Delta r_1$ (and $\Delta r_2$)  as a function of $1/L$;
  ; we obtain $h^{m}_{\infty} = 1.03(1)$ [and $h^m_{\infty} = 1.04(1)$].
  In all panels,
  the vertical dashed line corresponds to the critical point $h = h_c$.}

  \label{figVarIsing} % Fig 1
\end{figure}

In order to gather more insight about the behavior of the configurations in the
proximity of the critical point, we compute the variance $\Delta r_1$ of the
distribution of the recorded distances between each configuration in the data
set and its NN; our results for this quantity for various system sizes are
shown in Fig.~\ref{figVarIsing}(a) as a function of the transverse field. The
variance $\Delta r_1$ displays a peak at a size-dependent position $h^m(L)$,
with the distribution $f(r_1)$ correspondingly displaying significant
broadening when approaching the critical point [Fig.~\ref{figVarIsing}(b)]. The
$h^m(L)$ are not necessarily identical to the $h^*(L)$, but display the same
behavior, gradually approaching the critical point for increasing system size.
Performing a linear fit as in the case of the $I_d$ minimum position [see
Fig.~\ref{figVarIsing}(c)], we obtain the extrapolated value $h^m(\infty) =
1.03(1)$. The variance of the distribution for the distance of each
configuration to its NNN in the sampled data set (not shown) displays an
essentially identical behavior, and a fit performed in the same fashion as
above yields an extrapolated value $h_{\infty}^m = 1.04(1)$ for the peak
position in the thermodynamic limit.

Remarkably, we observe that the singular features associated with both the
$I_d$ and $\Delta r_1$ (i.e., the minimum in the former and the peak of the
latter) shift with $L$ from the ordered phase towards the critical point, in
the same fashion as the finite-size peak of $\chi_x$. These results suggest a
relation between these observables and the correlations associated with $S^x$
degrees of freedom, which encompass both spatial and imaginary-time degrees of
freedom, and are therefore deeply connected to the quantum nature of the
problem.  The relation between the latter and the behavior of the
data-mining-inspired quantities is also immediately evident from a direct
comparison with the classical counterpart of the quantum problem investigated
here, i.e., the paramagnetic-ferromagnetic transition in a 2D Ising model,
examined in \cite{Mendes2020}. In this case, where quantum fluctuations are
absent, the shift of the singular feature of the $I_d$ behaves like the one of
the ``diagonal'' (i.e., classical) susceptibility $\chi_z \equiv L \beta \left(
\ave{m_z^2} - \ave{\left| m_z \right|}^2 \right)$ (i.e., it occurs at finite
sizes within the paramagnetic phase, unlike in the quantum case, converging to
the critical point for $L \to \infty$).

\begin{figure}[t]
  \begin{center}
    {\centering\resizebox*{8.8cm}{!}{\includegraphics*{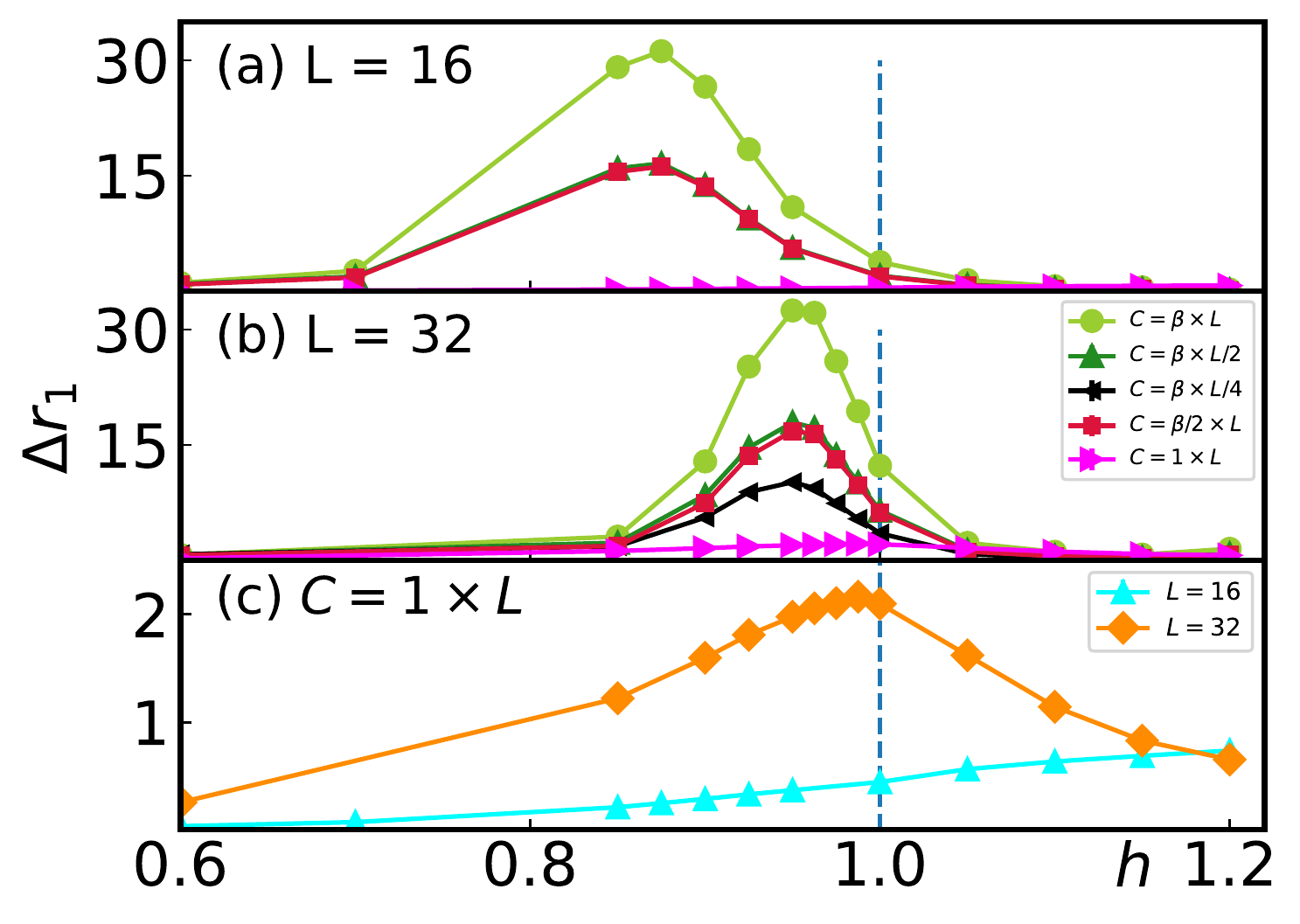}}}
  \end{center}

  \caption{\textit{Quantum Ising model (statistics of NN distances for
  different system partitions).} Panel (a): variance of the distribution
  function of the NN distances as a function of the transverse field for the
  complete data set of $L = 16$ sites $\times$ $M' = \beta = 512$ slices (green
  circles), a spatial half-partition with the first $L/2$ sites and $\beta$
  slices (dark green triangles), a temporal half-partition of the system with
  all $L$ sites but where only one every two of the original $M'$ slices is
  considered (red squares), and a data set composed by a single slice for the
  complete system of $L$ sites (purple triangles). Panel (b): same as panel (a)
  for $L = 32$, with the addition of the data set corresponding to a
  quarter-chain partition (i.e., the first $L/4 = 8$ sites) and all slices.
  Panel (c): magnification of the curves for the NN distance variance for
  single-slice data sets for $L = 16, 32$. In all panels, the vertical dashed
  line corresponds to the critical point $h = h_c$.}

  \label{fig.partitions}
\end{figure}

In order to understand the role of spatial and imaginary-time degrees of
freedom of the path integral representation in determining the behavior of
$I_d$-related features, such as the variance peak of the NN distance
distribution, we compare the behavior of the latter observable computed for
different spatial and temporal partitions of the system. Our results are
displayed in Fig. \ref{fig.partitions}(a-b) for $L = 16$ and $L = 32$,
respectively.

Regardless of the subset of degrees of freedom considered, in proximity of the
critical point the observables display the same qualitative characteristics as
their counterparts for the complete system, i.e., a peak for a size-dependent
value of the transverse field. However, the details of such features, such as
the height and position of the peak, depend on the number of degrees of freedom
(sites $\times$ slices) considered: for instance, halving the number of degrees
of freedom (either in space, by considering a half-chain partition, or in
imaginary time, by considering only one slice every consecutive two) results in
a roughly halved peak height, and the features are likewise much weaker in the
case of single-slice calculations [but still present, see
Fig.~\ref{fig.partitions}(c)]. This behavior shows essentially the same
characteristics regardless of whether the ``excised'' degrees of freedom are
spatial or temporal, suggesting an essentially equivalent role of the two in
the calculation of $I_d$-related features.

Direct analysis of the features of $I_d$-related observables such as those
displayed in, e.g., Fig.~\ref{figVarIsing}, points out that the strength of the
$I_d$-related features (e.g., the height of the NN variance distribution peak)
does not increase indefinitely with the addition of more degrees of freedom,
but eventually reaches a size-independent saturation value corresponding to a
complete system once a high enough number of temporal degrees of freedom is
considered. It is also evident that, even for equivalent and large enough
number of slices to reach the saturation threshold mentioned above, an
$L$-sized partition of a larger system displays different features with respect
to a full system of $L$ sites (as can be seen comparing the results for the $8
\times M' = 512$ partition in Fig.~\ref{fig.partitions} with those of the full
$L = 8$ system, and the same number of slices, in Fig.~\ref{figVarIsing}).

\subsection{Two-dimensional dimerized models} \label{sec.2ddm}

\begin{figure}[t]
\begin{center}
{\centering\resizebox*{8.6cm}{!}{\includegraphics*{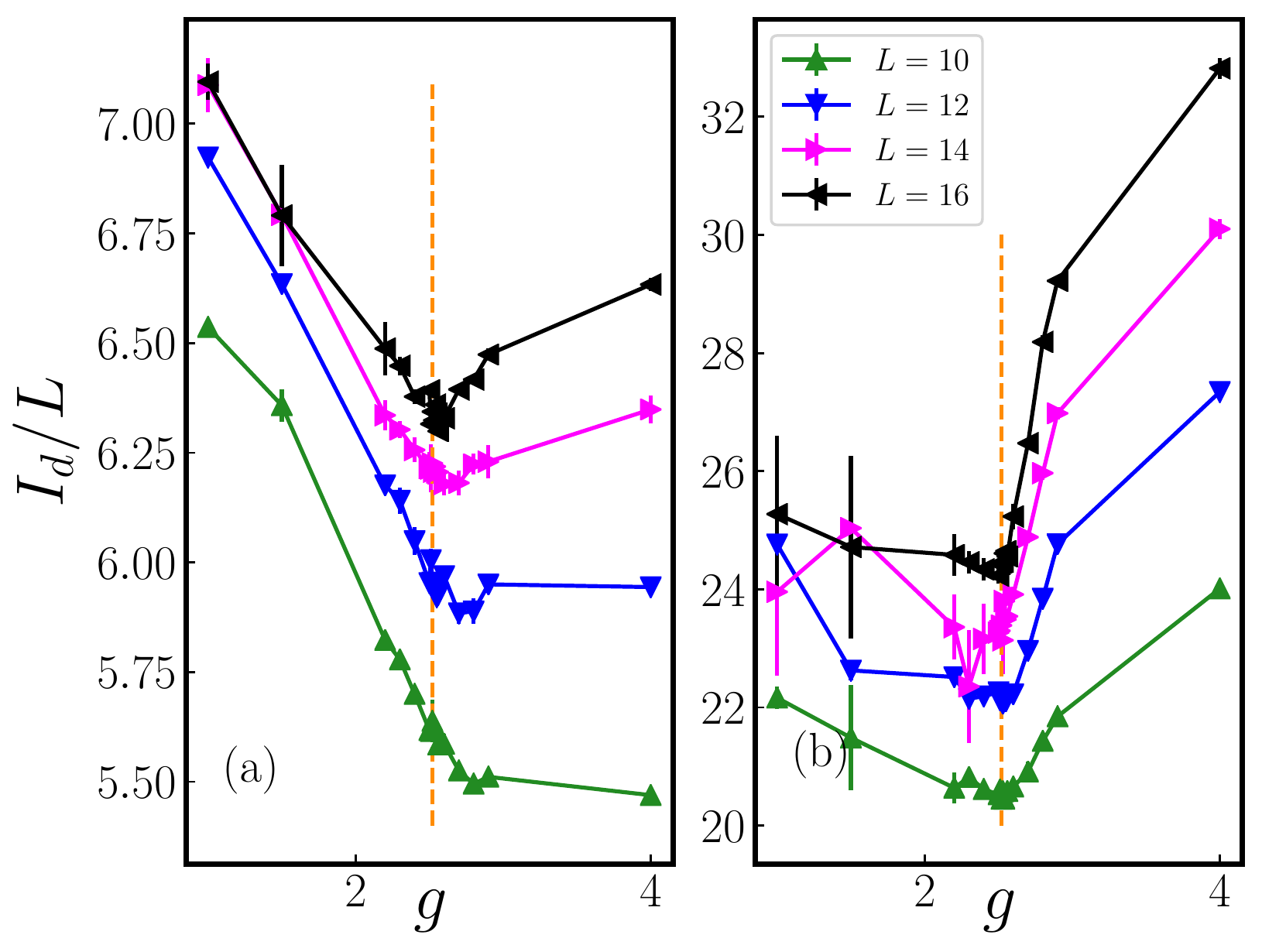}}}
\end{center}
\caption{\textit{Heisenberg bilayer (intrinsic dimension).} The graph shows the
  intrinsic dimension $I_d/L$ as a function of $g$ for different values of $L$.
  In panel (a) we consider data sets containing a single slice, while in panel
  (b) we consider ones with $\beta$ slices. For all results displayed here
  $\beta = L$. In all panels, the vertical dashed line corresponds to the
  critical point $g = g_c$.}
\label{figIDBilayer} % Fig 1
\end{figure}

\begin{figure*}[t]
\begin{center}
{\centering\resizebox*{9.2cm}{!}{\includegraphics*{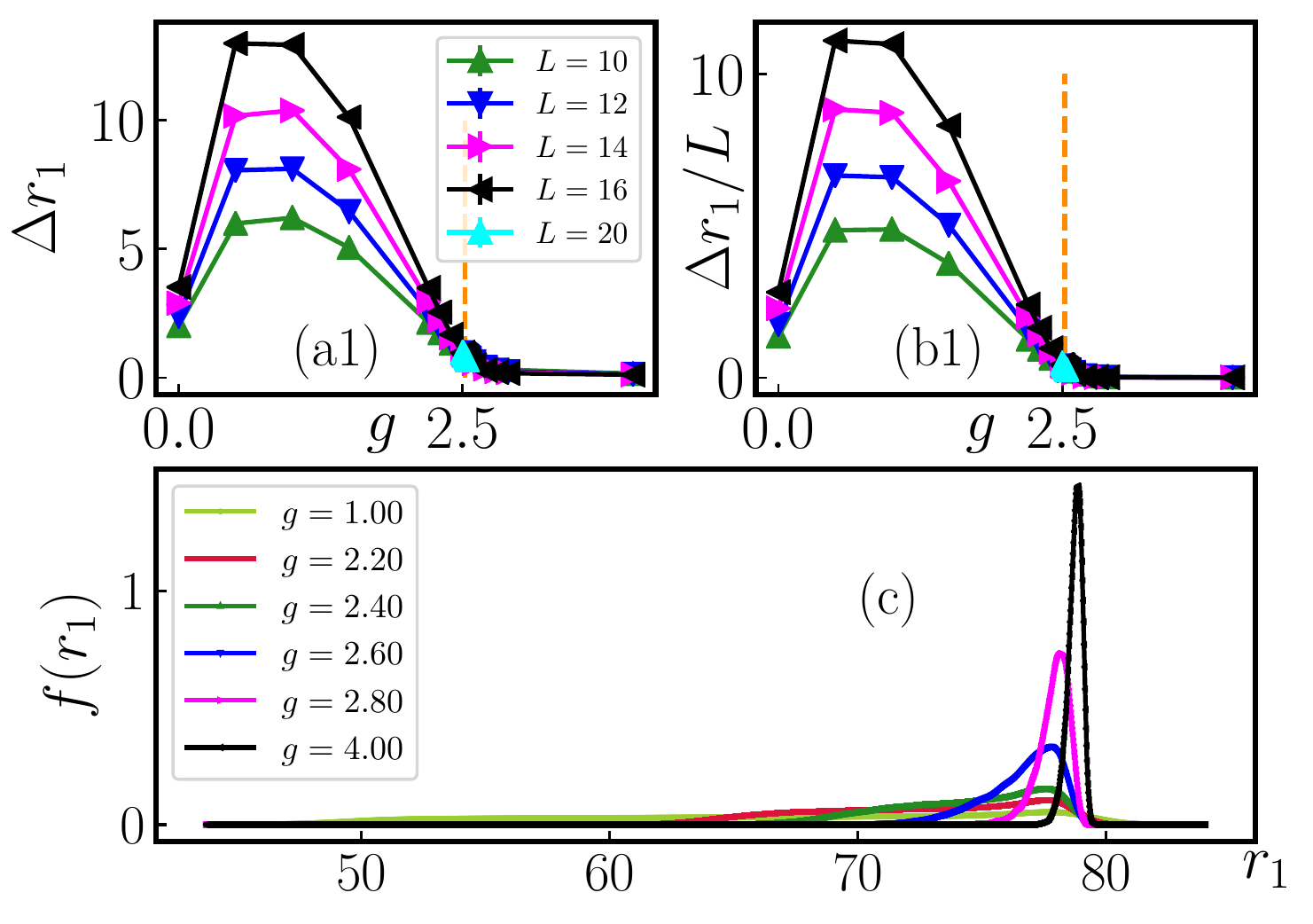}}}
{\centering\resizebox*{8.6cm}{!}{\includegraphics*{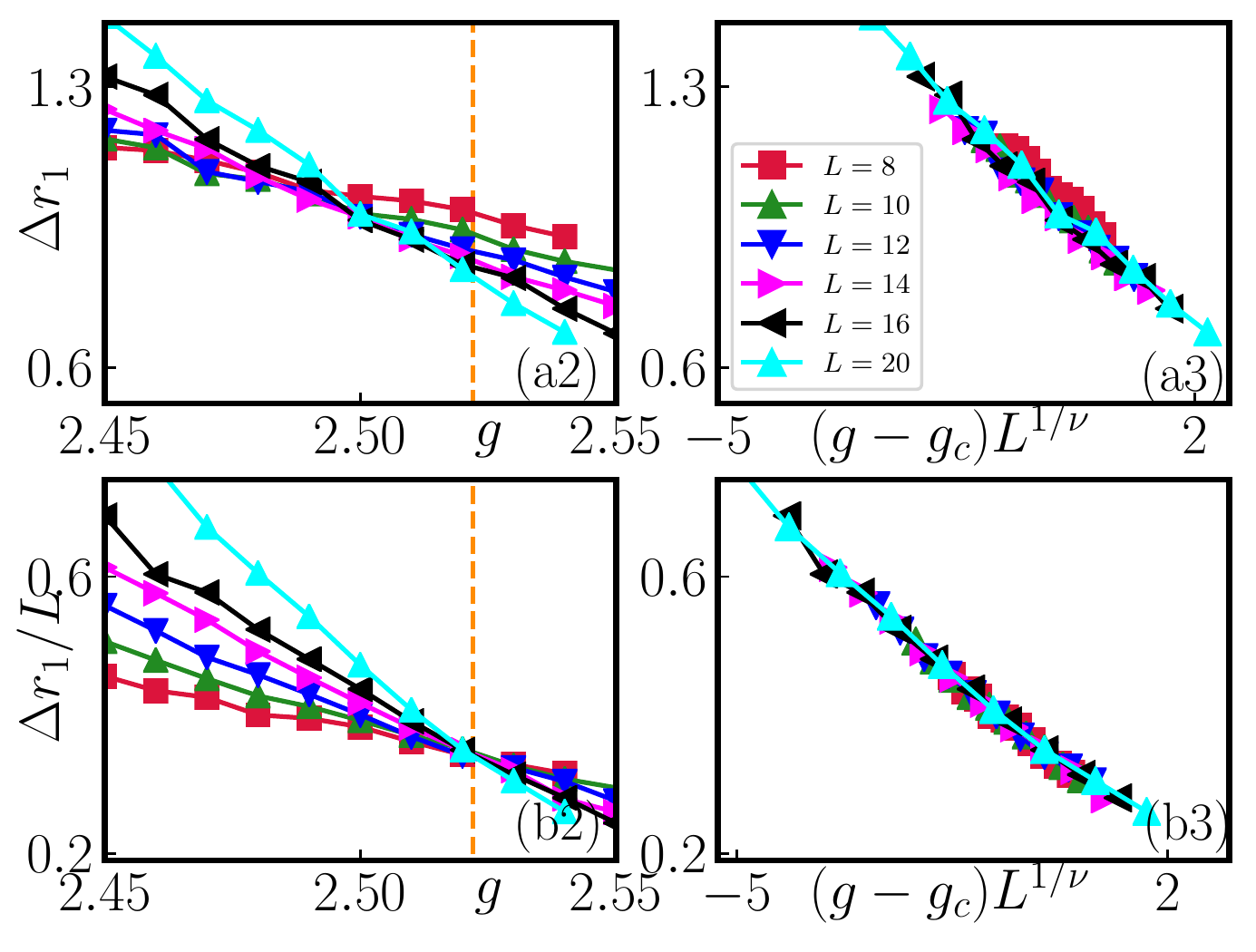}}}
\end{center}
\caption{\textit{Heisenberg bilayer (statistics of NN distances).} Panels
  (a1) and (a2) show the variance of the distribution function of the NN
  distances $f(r_1)$, $\Delta r_1$, as a function of $g$ for single- and
  $\beta$-slices data sets, respectively; for the $\beta$-slices data set, we
  consider $\Delta r_1/L$ (see text). In panel (c), we show an example of
  $f(r_1)$ for the $\beta$-slices data set and $L = 10$. In panels (a2) and
  (b2) we present the blow-up of (a1) and (b1), respectively, in the region
  close to $g_c$. Finally, we show the data collapse of the results in panels
  (a3) and (b3). The value of $\nu$ is in agreement with the expected 3D
  Heisenberg universality class. For all results displayed here $\beta = L$. In
  all panels, the vertical dashed line corresponds to the critical point $g =
  g_c$.}
\label{figVarBilayer} % Fig 1
\end{figure*}

We now consider the 2D dimerized Heisenberg bilayer model in Eq.
\eqref{eq.ham-bilayer}. This Hamiltonian describes an AFM-paramagnetic
transition belonging to the same universality class of the three-dimensional
$O(3)$ Heisenberg model, see Sec. \ref{IIb}.

Our simulations are performed using the SSE algorithm at inverse temperature $\beta = L$, an appropriate value for the investigation of the ground-state regime \cite{Sandvik1991,Sandvik1997}. For all the results discussed here, we consider data sets containing $N_r = 5 \times 10^4$ configurations. Furthermore, we consider configurations containing (i) single or (ii) a set of $M' = \beta$ equally spaced slices.
In order to ensure that the configurations belonging to the data set are uncorrelated, we computed the $I_d$ and $\Delta r_1$ of data sets generated by sampling configuration separated by $n_{AC}$ Monte Carlo steps, analyzing the dependence of the obtained estimates on the value of the latter. Our analysis resulted in the observation that, while our results depended strongly on the value of $n_{AC}$ for small inter-configuration distance, the values of $I_d$ and $\Delta r_1$ stabilized for $n_{AC} \gtrsim 10$, as expected once the decorrelated regime is reached. All results obtained via the SSE algorithm discussed here have been obtained following the procedure outlined above with $n_{AC} \ge 100$, ensuring the decorrelation of the analyzed configurations.

First, let us consider the behavior of the $I_d$. As in the case of the quantum
Ising transition, the $I_d$ features a minimum close to the critical point
$g_c$, which interestingly appears both for (i) single-slice [see Fig.
\ref{figIDBilayer}(a)] and (ii) $\beta$-slices data sets [see Fig.
\ref{figIDBilayer}(b)]. We note that the $I_d$ minimum position slightly
changes when more slices are considered in the data set. In particular, for
$\beta$-slices data sets and $L = 16$, the minimum of the $I_d$ is located
within $1\%$ of the QMC estimation of $g_c$ \cite{Sandvik2006}.

The analysis of the distribution of NN distances, $f(r_1)$, provides further
insight into the data structure emerging in the vicinity of $g_c$. The first
striking observation is the non-monotonic behavior of the variance of $f(r_1)$,
$\Delta r_1$, for $g < g_c$, which displays a peak for $g \approx 0.5$
(interestingly, this is the same behavior observed for the AFM order parameter
\cite{Sandvik1994, Mendes2017}). This feature occurs for both the single- and
$\beta$-slices data sets, which indicates that it is independent of the number
of slices of the configurations; see Fig. \ref{figVarBilayer}(a1-b1). It is
worth mentioning that this behavior is qualitatively different from the one
observed for $\Delta r_1$ for the quantum Ising transition: in particular, here
the position of the variance peak is (roughly) size-independent, and does not
shift towards to the critical point. This difference is related to the
underlying symmetries of these models - the SU($2$) of the Heisenberg bilayer
and the $\mathbb{Z}_2$ of the quantum Ising model - and the corresponding
symmetry-broken phases. Before explaining these results (see Sec.
\ref{sec:sym}), which are of genuine quantum mechanical nature and pointing out
the sharp difference between quantum and classical data sets, let us discuss
the behavior of $\Delta r_1$ in the vicinity of $g_c$.

Indeed, we observe that $\Delta r_1$ ($\Delta r_1/L$) is (almost) a
$L$-independent quantity close to $g_c$ when single-slice ($\beta$-slices) data
sets are considered. The transition is then accurately identified by the
crossing point of $\Delta r_1$ curves associated to different values of $L$, as
illustrated in Fig. \ref{figVarBilayer}(a2-b2). In addition, Fig.
\ref{figVarBilayer}(a3-b3) show that our results are well described by the FSS
\textit{ansatz} $\Delta r_1$ ($\Delta r_1/L$) $ = f[(g-g_c)L^{1/\nu}] $, where
$\nu$ is the critical exponent associated to the divergence of the correlation
length, for single-slice ($\beta$-slices) data sets. Our results are $g_c =
2.50(2)$ and $\nu = 0.71(2)$ ($g_c = 2.52(1)$ and $\nu = 0.68(2)$) for single
slices ($\beta$-slices) data sets, differing of less than $1\%$ ($5\%$) from
accurate estimations of these quantities based on FSS of physical observables
\cite{Sandvik2006}.

Finally, it is worth mentioning that such a scaling behavior in the vicinity of
2D quantum critical points is also displayed by physical quantities including,
e.g., the Binder moment ratios and the (rescaled) spin stiffness $L \rho_s$
\cite{Sandvik2006}. Accurate estimations of critical points and exponents can
be obtained with the same strategy adopted here, i.e, the detection of crossing
points of results for different values of $L$. For example, one can determine
the crossing points $g_c(L)$ for results corresponding to system sizes $L$ and
$2L$, and then use FSS techniques to establish the value of $g_c(L \to
\infty)$, which constitutes an estimate for the critical point. An interesting
observation is that that the $g_c(L)$ associated to $\beta$-slices data sets
converge faster to $g_c(L \to \infty)$ than the results for single-slice data
sets, see Fig.  \ref{figVarBilayer}(a2-b2). This is in line with what is observed
in conventional FSS of physical quantities: i.e, the $g_c(L)$ associated with
different observables exhibit different convergence to the $L \to \infty$
limit, as a consequence of subleading corrections of scaling functions. In this
case, the crossing points associated with the spin stiffness exhibit the most
rapid convergence to the thermodynamic limit \cite{Sandvik2006}. Interestingly,
$\rho_s$ is a non-equal-time quantity that depends on the full space-time
structure of the path integral, which the $\beta$-slices data sets represent in
the most faithful way.

To get further evidence for the conclusions described in this section, we also
consider the Heisenberg columnar dimer model (see Appendix \ref{appDi}). This
Hamiltonian also describes a AFM-paramagnetic transition belonging to the same
universality class of the three-dimensional $O(3)$ Heisenberg model. The
results for the $I_d$ and $\Delta r_1$ are equivalent to the ones described
above (see Fig. \ref{FigDi} in Appendix \ref{appDi}), emphasizing how generic
features of data structures are solely determined by the universal properties
of the underlying quantum critical point.

\section{BKT transition}
\label{sec:result2}

\begin{figure}[t]
\begin{center}
{\centering\resizebox*{8.8cm}{!}{\includegraphics*{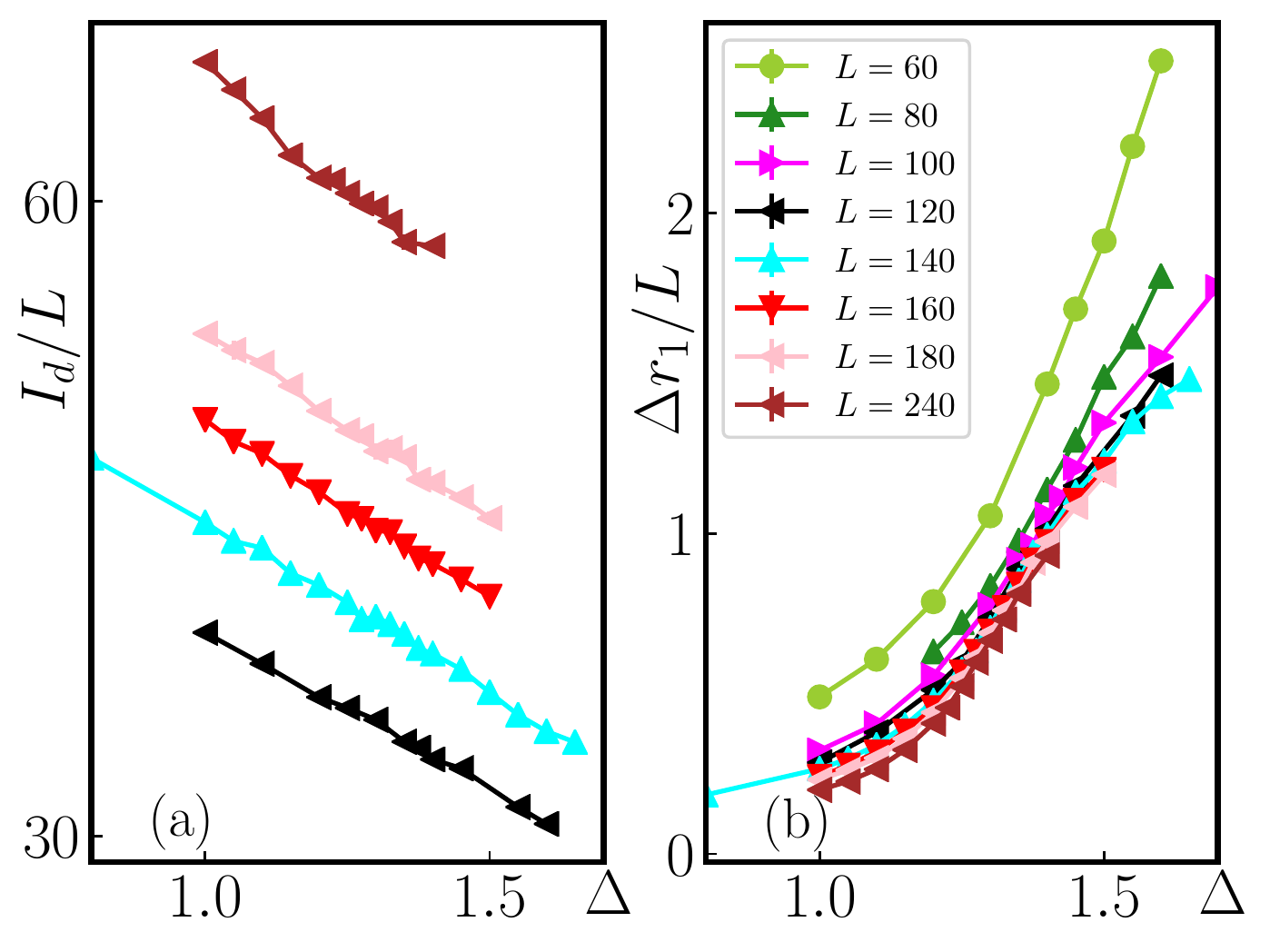}}}
\end{center}
\caption{\textit{BKT transition - one-dimensional XXZ model. Results for
  single-slice data sets.} Panel (a) and (b) display the $I_d$ and $\Delta
  r_1$, respectively, as a function of $\Delta$. For all results shown here
  $\beta = L$.}
\label{figxxzSingle} % Fig 1
\end{figure}

\begin{figure}[t]
\begin{center}
{\centering\resizebox*{8.8cm}{!}{\includegraphics*{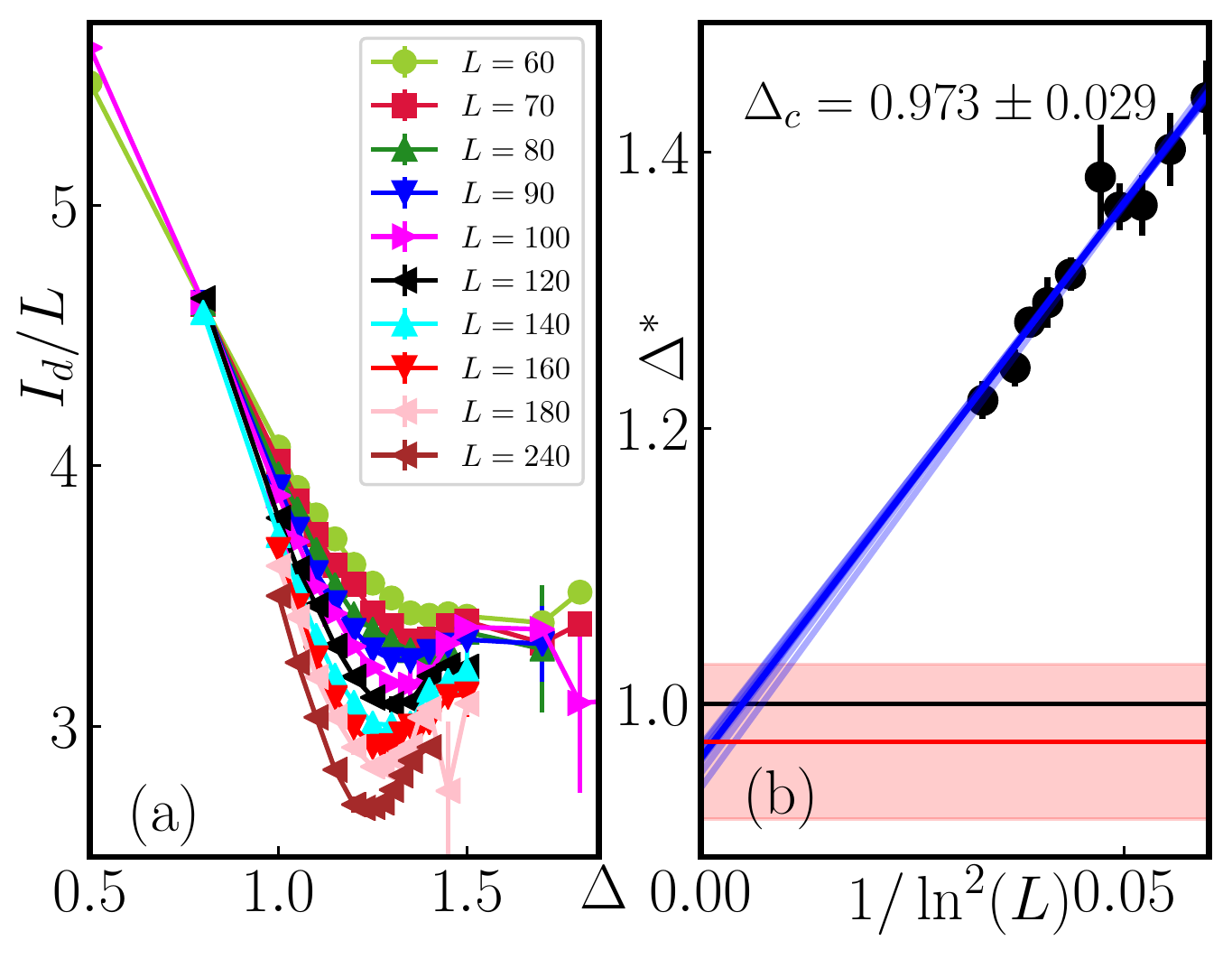}}}
\end{center}
\caption{\textit{BKT transition - one-dimensional XXZ model. Results for the
  $\beta$-slices data sets.} Panel (a) shows the intrinsic dimension, $I_d/L$,
  as a function of $\Delta$ for different values of $L$. In panel (b) we
  consider the finite-size scaling of the positions $\Delta^*(L)$ of the $I_d$
  minima. The (blue) lines are line fittings performed with different sets of points, and the red horizontal line corresponds to the averaged $\Delta_c$ computed with such fittings. 
  For all results displayed here $\beta = L$.}
\label{figIDxxz} % Fig 1
\end{figure}

\begin{figure}[t]
\begin{center}
{\centering\resizebox*{9.2cm}{!}{\includegraphics*{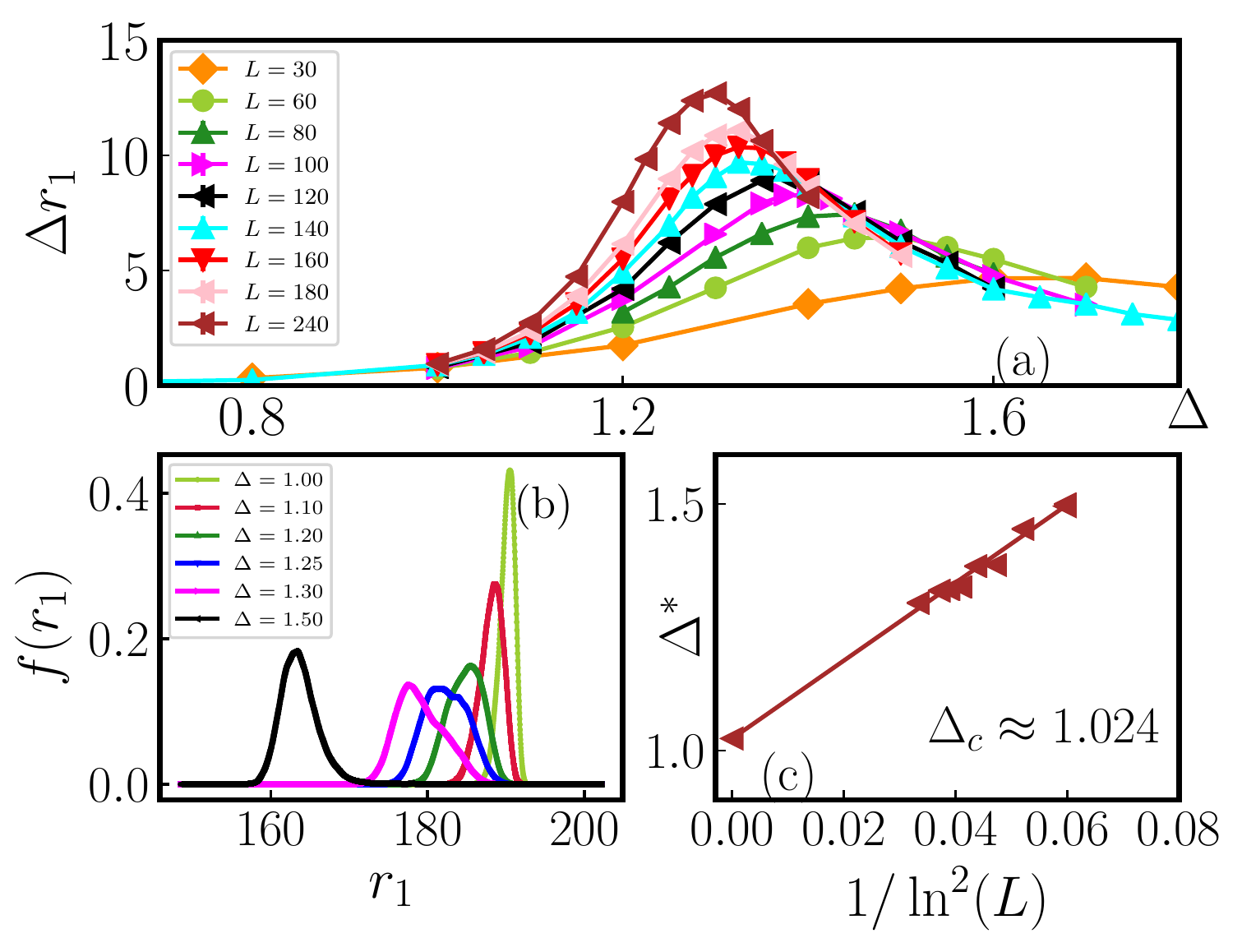}}}
\end{center}
\caption{\textit{BKT transition - one-dimensional XXZ model. Results for the
  $\beta$-slices data sets.}  Panel (a) shows the variance of the distribution
  function of the NN distances $f(r_1)$, $\Delta r_1$, as a function of
  $\Delta$ for different values of $L$. In panel (b), we show an example of
  $f(r_1)$ for $L = 140$ and different values of $\Delta$. Finally, in panel
  (c), we present the finite-size scaling of the positions $\Delta^m(L)$ of the
  maxima of $\Delta r_1$. For all results displayed here $\beta = L$.}
\label{figVarxxz} % Fig 1
\end{figure}

In this section we consider the BKT transition described by the one-dimensional
spin-$1/2$ XXZ model in Eq. \eqref{eq.ham-xxz}. Differently from the cases
discussed above, BKT quantum critical points (QCPs) are characterized by
physical quantities associated to global properties of the full path integral
configurations \cite{Gogolin2004, Giamarchi2004, sachdev_2000}. For instance,
they are conventionally described by the spin stiffness, which is related to
fluctuations of a topological property of path integral configurations (namely,
the winding number). Nonlocal quantum-information quantities~\cite{Karyn2012},
and spectral properties~\cite{Rigol2011, Dalmonte2015} are also used to pin
down BKT QCPs.

The nature of BKT QCPs hints that successful unsupervised learning of such
transitions relies (i) on the one hand, on defining a proper data set, that
encompasses BKT topological properties, and (ii) on the other hand, on
analyzing data set features that can reveal such global properties. Before
discussing our results, let us mention that for the classical 2D XY model the
$I_d$ exhibits signatures of the BKT phase transition \cite{Mendes2020}.
Furthermore, dimensional reduction methods based on diffusion maps can detect
the topological clustering structure of such classical data sets
\cite{Scheurer2019, Wei2021}. Diffusion maps have also been used to reveal
other topological properties (not related to BKT transitions) \cite{Yang2020,
Lidiak2020, Nori2020}.

Let us now discuss how one can detect a BKT QCP with both the $I_d$ and $\Delta
r_1$ by defining suitable path integral data sets. We employ the directed-loop
SSE method to sample the path integral configurations, following the same
protocol outlined in Sec. \ref{sec.2ddm}.

First, we consider the results associated with data sets containing
single-slice configurations. In this case, we observe no particular feature in
the behavior of either the $I_d$ or $\Delta r_1$ close to $\Delta_c = 1$, see
Fig. \ref{figxxzSingle}. Subsequently, we consider data sets formed by
$\beta$-slices data sets. A fundamental technical aspect is that we retrieve
slices separated by an interval of $\delta \tau \approx N_s$ in the SSE
``imaginary-time'' direction. Remarkably, in this case we observe that both the
$I_d$ and $\Delta r_1$ exhibit singular features in the vicinity of $\Delta_c$,
see Figs. \ref{figIDxxz} and \ref{figVarxxz}.

More specifically, the $I_d$ features minima at size-dependent positions
$\Delta^*(L)$ in the vicinity of $\Delta_c$. By performing a finite-size
scaling of the minimum positions we obtain an estimate for the critical
$\Delta_c$, see Fig. \ref{figIDxxz}(b). Furthermore, we consider the data
collapse of these results according to the FSS \textit{ansatz} $I_d = L^{1
-\omega} f(\xi/L)$, where the correlation length diverges as $\xi \sim
\exp{(b/\sqrt{\Delta - \Delta_c})}$ [see Fig. \ref{fig1}(c)]. The value
obtained for $\Delta_c$ via this procedure is in agreement with the exact one.

The statistics of NN distances $r_1$ also reveal the BKT quantum critical
point. In particular, we note that the variance $\Delta r_1$ of the
distribution $f(r_1)$ exhibits a peak at size-dependent points $\Delta^m(L)$ in
the vicinity of $\Delta_c$. The $\Delta^m(L)$ shifts towards $\Delta_c$ as $L
\to \infty$, and their scaling to the thermodynamic limit allows to obtain an
estimation of $\Delta_c$, see Fig. \ref{figVarxxz}(b).

We also present the data collapse of our results using the FSS \textit{ansatz}
$\Delta r_1 = L^{-\alpha} \tilde{f}(\xi/L)$ (Fig. \ref{fig2}). The quality of
the collapse and the value obtained for $\Delta_c$ provides further numerical
evidence that $\Delta r_1$ exhibits the universal scaling behavior
characteristic of BKT QCPs.

Before concluding this section, let us mention about the influence of $\beta$ on our results. In our numerical simulations, we have found that as long as the value of $\beta$ is large enough to guarantee convergence to the ground state, the results for single-slice data sets are not affected by the choice of $\beta$.
Oppositely, for the case of $\beta$-slices data sets, the values of $I_d$ and $\Delta r_1$ does depend on $\beta$, because the dimension of the embedding space changes with the latter; however, we observe that all the features associated to the phase transition of both $I_d$ (i.e., the minimum) and $\Delta r_1$ (i.e., the peak) do not change as $\beta$ increases (again, as long as $\beta $ is large enough to guarantee convergence to the ground state).

\subsection{Discussion of the results}

Summing up the results presented so far, we observe that both $I_d$ and $\Delta r_1$ exhibit singular features \footnote{It is important to mention that strictly speaking the term ``singular features'' mean that a physical quantity (or some of its derivatives) diverges in the thermodynamic limit.
In this work, although we do not show that $I_d$ and $\Delta r_1$ (or some of their derivatives) exhibit divergences, we employ the term ``singular features''. This is motivated by our observation that such quantities exhibit scaling behavior close to quantum critical points. As occurs with singular physical quantities, this indicates that the correlation length dictates the behavior of   $I_d$ and $\Delta r_1$.} in the vicinity of both symmetry-breaking and BKT phase transitions. 

Although we focus on paradigmatic models, we will now present arguments supporting the idea that these results are generic for the quantum data sets defined in Sec. \ref{quantumdataset}, and thus also apply for similar QCPs as the ones considered here.
In previous work, some of us presented heuristic arguments to explain the relationship between the ID and the physical correlation length in classical phase transitions \cite{Mendes2020}.
The basic idea lies in the hypothesis that the neighboring distances [see Eq. \eqref{disE}] $r_1$ and $r_2$ are related to many-body correlations of the physical degrees of freedom of the system. This hypothesis can be straightforwardly applied to the quantum case for the specific setting in which the dynamical critical exponent satisfies $z = 1$. In these cases, the intrinsic dimension is directly related to arbitrary rank correlation functions - and thus, shall display characteristic scaling behavior at the transition.

It is important to note that these arguments imply that 
$I_d$ and $\Delta r_1$ depend on correlations arising from both spatial and imaginary-time degrees of freedom. 
Our results indeed suggest that this is the case for $\beta$-slices data sets. For example, we mention the scaling of $I_d$ and $\Delta r_1$ for the TFIM shown in Figs. \ref{figIDIsing} and \ref{figVarIsing}, which is analogous to physical quantities encompassing space-time degrees of freedom. Furthermore, in the case of BKT, we point out that single-slice data sets (encompassing solely spatial degrees of freedom) are insufficient to determine whether a transition exists at all, a fact which agrees with the topological nature of quantum BKT transitions.

\section{Quantum data sets, symmetries, and symmetry-broken phases}
\label{sec:sym}

So far we focused on the results emerging in the vicinity of different QCPs.
We now argue that some of the data set features analyzed here can also reveal
fundamental properties of symmetry-broken phases. In particular, let us review
the behavior of $\Delta r_1$ in the different ordered phases encountered in
this work: namely, (i) the $\mathbb{Z}_2$ ferromagnet described by the quantum
Ising model, (ii) the Luttinger liquid and the $\mathbb{Z}_2$ antiferromagnet
displayed in the 1D XXZ model, and (iii) the SU($2$) antiferromagnet described
by the Heisenberg Bilayer.

\begin{figure}[t]
\begin{center}
{\centering\resizebox*{9.1cm}{!}{\includegraphics*{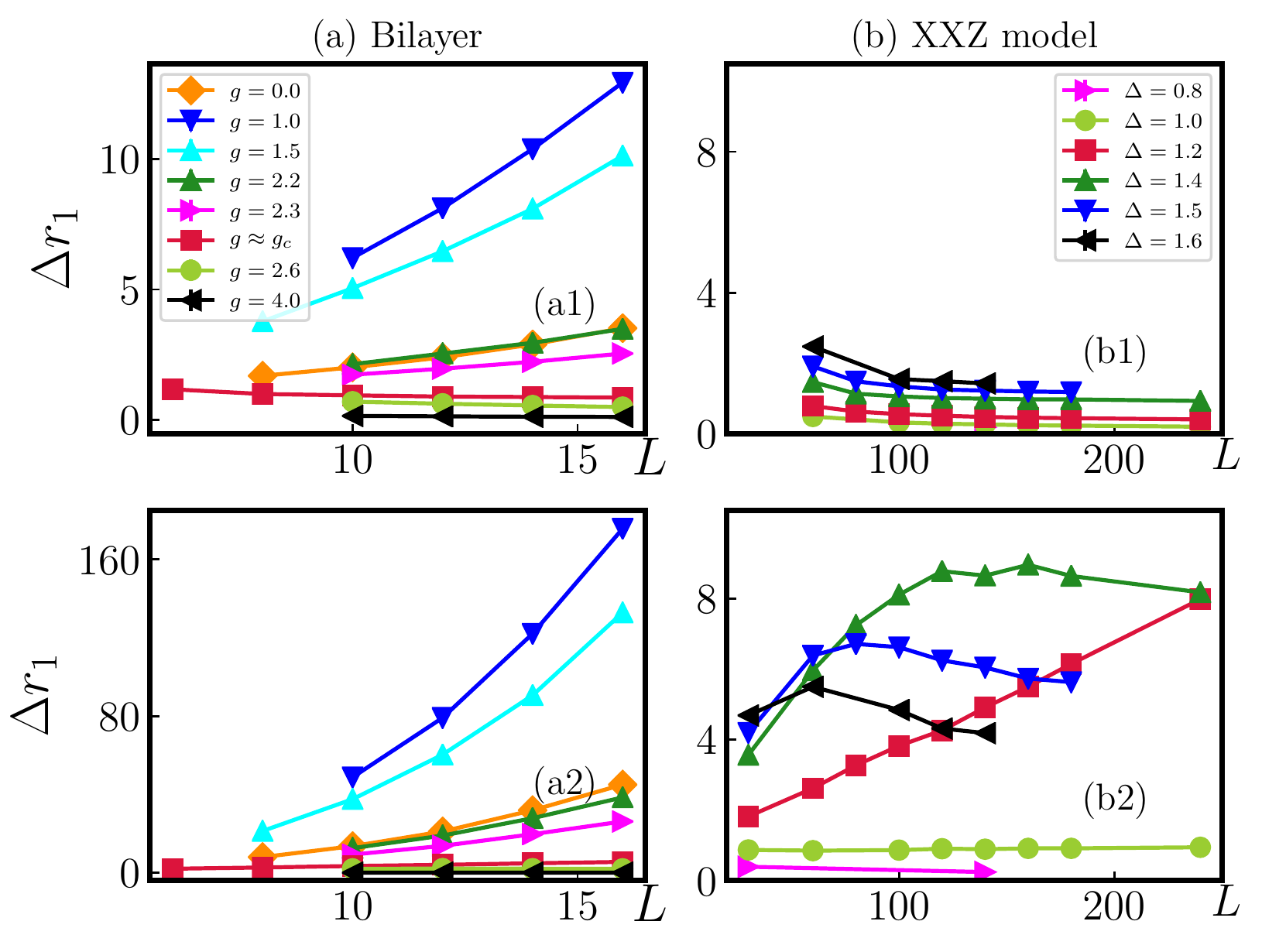}}}
\end{center}
  \caption{\textit{Size dependence of NN distance variance.} Variance of the
  distribution function $f(r_1)$, $\Delta r_1$, as function of the system size
  $L$.  Panels (a) and (b) show our results for the 2D Heisenberg Bilayer and
  1D XXZ models, respectively. In panels (a1) and (b1) we consider data sets
  generated by single-slice configurations, while in (a2) and (b2) we consider
  configurations containing $\beta$ slices. For all results displayed here
  $\beta = L$.}
\label{figVarL} % Fig 2
\end{figure}

Away from quantum critical points, we can summarize our results as follows:
apart from the SU($2$) AFM case, $\Delta r_1$ is always an intensive (or weakly
dependent on $L$) quantity, in both ordered and disordered phases. To
illustrate this, in Fig.~\ref{figVarL}(b1-b2) we depict the scaling of the NN
distance variance versus system size for the XXZ model. In the case of
single-slice data sets (upper panel), the variance does not grow with $L$. In
the $\beta$-slices case, the variance does not grow in the gapless phase, while
in the AF phase it grows until it reaches the correlation length, and then
starts decreasing (likely approaching a constant).

Opposite to this, in the SU($2$) AFM phase $\Delta r_1$ is an extensive
quantity, i.e, $ \Delta r_1 \sim L$ (or $ \Delta r_1/L \sim L$ for
$\beta$-slices data sets). This behavior is depicted in
Fig.~\ref{figVarL}(a1-a2). Below we argue that this result is related to the
non-Abelian nature of the SU($2$)-symmetry-broken AFM phase, and reflects
fundamental aspects of the quantum data sets.

The latter considered here are labeled by the value of commuting local observables (i.e., $z$ components of spin-$1/2$ degrees of freedom), while the full quantum state is characterized by the expectation values of non-commuting observables as well. An important aspect to consider is that one cannot measure simultaneously more than one local spin component (i.e., $S^x_i$, $S^y_i$, and $S^z_i$) in quantum data sets.

In order to show how this \textit{quantum aspect} affects the results for
$\Delta r_1$, we compare its scaling in the SU($2$) Heisenberg bilayer [see
Fig. \ref{figVarBilayer}] with its classical counterpart, i.e., the  classical
O($3$) model. In the latter case, there is no problem in retaining a fully
invariant SU(2) description, and we can define data sets formed by either (a)
configurations $\vec{X} = (\vec{S}_1,...,\vec{S}_{N_s}$), where $\vec{S}_i =
(S^x_i,S^y_i,S^z_i)$, and (b) configurations defined just by the $z$-components
of the spins [i.e., $\vec{X} = (S^z_1,...,S^z_{N_s})$], as in the quantum case.

\begin{figure}[t]
\begin{center}
{\centering\resizebox*{9.0cm}{!}{\includegraphics*{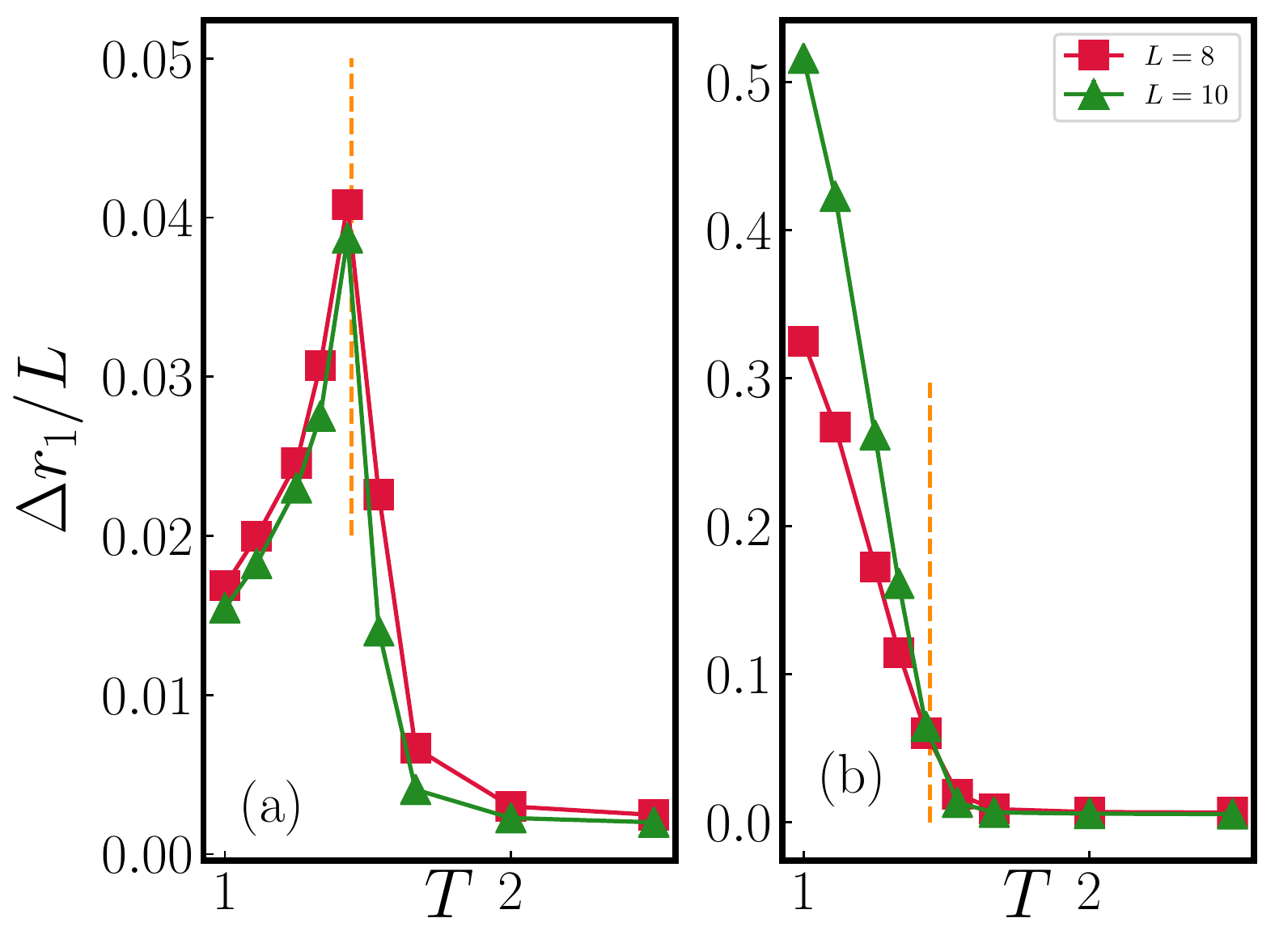}}}
\end{center}
\caption{\textit{Classical Heisenberg model.} Panels (a) and (b) show the
  variance of the distribution function of the NN distances $f(r_1)$, $\Delta
  r_1$, as a function of $T$. In panel (a) we consider data sets generated by
  all $3$ spin components (i.e., $S^x$,$S^y$ and $S^z$), while in (b) we only
  consider the $S^z$ component (see text). In both panels, the vertical dashed
  line corresponds to the critical point, $T_c \approx 1.443$ \cite{Holm1993}.}
\label{figHeis} % Fig 1
\end{figure}

In Fig. \ref{figHeis} we consider the temperature dependence of $\Delta r_1$ for the the classical O($3$) model. While for the data sets (a) $\Delta r_1$ exhibits a peak in the vicinity of the critical temperature $T_c$, for the data sets (b) $\Delta r_1$ sharply increases in the ordered phases (i.e., $T < T_c$). This reflects exactly what happens in the quantum case: when the full SU(2) symmetry is not resolved, the structure of the manifold changes drastically in symmetry-broken phases, and the variance of the distribution of distances increases extensively with system size. Note that the overall scale of $\Delta r_1$ is also very different: while the data sets (a) are embedded in a manifold that is three times as large (in terms of number of dimensions) as the (b) ones, $\Delta r_1$ at $L=10$ is an order of magnitude smaller in the symmetry-broken phase.

A qualitative explanation of the effect of SU(2) symmetry goes as follows: if a
symmetry is not fully resolved, it is not possible to identify apparently
different states as representatives of the same original state up to symmetric
transformations, leading to artificially generated non-local correlations in
the data sets, and to an enormously increased variance of distances. This
effect becomes particularly evident in the case of symmetry breaking: the
reason here is that clusters corresponding to different symmetry-broken regions
are well separated, something that is not expected to happen in either critical
or paramagnetic phases.

We can compare this picture with what happens when an Abelian symmetry is
broken. The results for $\Delta r_1$ within the Ising FM phase (see Fig.
\ref{figVarIsing}) can also be compared with its classical counterpart, i.e.,
the two-dimensional Ising model (see ref. \cite{Mendes2020}). In this case,
the classical data sets are defined exclusively in terms of $S^z$ variables,
and in the FM phase $\Delta r_1$ exhibit analogous behavior to its quantum
counterpart.

As a sanity check for the argument above, we can consider the critical point of
the 1D Heisenberg model. Indeed, this point does not exhibit the extensive
$\Delta r_1$ observed in its 2D counterpart. This difference shows that the
extensive behavior of $\Delta r_1$ is indeed characteristic of the quantum
fluctuations of the SU$(2)$-symmetry-broken phase, and not due the global
symmetry of the system.

Our results thus directly indicate that the presence of non-Abelian symmetries
can alter in a rather drastic manner the basic features of the PI manifold.
This increased complexity of the data structure may be the origin of a recent
set of observations in numerical studies using neural network \textit{ansatze}
as wave-functions. There, it was argued that neural network optimization may
suffer significantly in the absence of a fully resolved symmetry. Our results
are consistent with that observation, in that we provide evidence on why this
happens: the underlying embedding manifold has artificial correlations
introduced by the absence of symmetry resolution.

\section{Conclusions}
\label{sec:concl}

We have shown how features of the raw data structure of partition functions
reveal universal properties of both quantum phases and QCPs. The two key
elements underlying our approach are (i) the introduction of a properly defined
``raw data structure'', which is based on a proper treatment of space and
imaginary-time degrees of freedom of path integral (or equivalently,
stochastic-series-expansion) configurations generated by quantum Monte Carlo
simulations; and (ii) the investigation of generic features of quantum data
sets that are accessed without any dimensional reduction of the data set:
namely, the intrinsic dimension, $I_d$, and the variance of the distribution of
distances of NN configurations, $\Delta r_1$.

Our first key result is that both the $I_d$ and $\Delta r_1$ exhibit universal
scaling behavior in the vicinity of both symmetry-breaking (i.e., related to
the $\mathbb{Z}_2$ and the SU$(2)$ symmetries) and BKT QCPs. Data sets with a
single imaginary-time slice are already enough to reveal symmetry-breaking QCPs
at large enough system sizes. For the BKT transition, however, one needs to
consider configurations containing a proper set of multiple slices. These
results are traced back to the fact that while symmetry-breaking transitions
are described by local order parameters, BKT QCPs take place due to topological
changes of the full path integral configurations, and are associated with
nonlocal correlations (encompassing both space and imaginary-time degrees of
freedom). In this regard, our results elucidate the deep connection between
generic properties of data sets - associated with the statistics of neighboring
distances - and arbitrary-body correlations related to universal properties of
quantum phase transitions. We note that $I_d$ is in general more reliable than $\Delta r_1$, as the latter might be sensitive to inhomogeneities in the data structure.

The second key observation is that
the data structure of quantum partition functions simplifies (parametrically) as one approaches QCPs. Analogous conclusions are obtained for the conceptually simpler classical case~\cite{Mendes2020}. This finding has a clear implications related to the complexity of (equilibrium) quantum states. While the intrinsic dimension is not a rigorously defined measure of complexity, it still provides a very informative, quantitative tool to witness it: a higher-dimensional manifold will always require a larger number of coordinates to be described. Our results show that quantum criticality leads to a drastic reduction in complexity in critical phases, that reflects the fact that the PI is very constrained due to universality. This witness of complexity behaves in a manner that is antipodal to entanglement, as the latter is typically growing close to criticality. It would be interesting to understand whether these two, apparently different, ways of addressing complexity can be directly compared: one possibility in this direction would be to apply our method to the transfer matrix corresponding to a fixed-bond-dimension tensor network.

Our third key observation is that the raw data structure of quantum partition
functions is quantitatively affected by the spontaneous breaking of non-Abelian
symmetries. In particular, we observe that the $\Delta r_1$ of data sets
associated to the SU($2$) AFM phase exhibits dramatic differences compared to
the Abelian cases considered here. Indeed, in the former case $\Delta r_1$ is
an extensive quantity (i.e., $\Delta r_1 \sim L$), while in the latter $\Delta
r_1$ is almost independent of the system size (or even decreases with $L$).
The explanation for this result is compatible with a key aspect of the quantum
nature of the problem: namely, the SU($2$) non-Abelian symmetry cannot be fully
resolved by data sets defined in terms of local spin measurements.

Let us conclude this paper by mentioning some perspectives for our work, by
highlighting potential applications of the intrinsic dimension to other quantum
mechanical objects.

One possible application concerns experiments. The first type of data set
analyzed here (i.e., single-slice data sets) can be directly extracted via
\textit{in situ} imagining, at a similar cost to conventional correlation
functions~\cite{Brydges2019probing, Barredo, chiaro2020direct, Pagano_2020,
scholl2020programmable, ebadi2020quantum, Zeiher_2017, Pfau2021,
Madjarov_2020}. For example, one can consider the $I_d$ (or $\Delta r_1$) to
witness complexity in a manner that is considerably less expensive to analyze
than entanglement-related approaches, to detect phase transitions characterized
by symmetry breaking, or even to reveal the presence of SU($2$) symmetry
breaking from raw experimental data. We note that the number of realizations we
typically consider here are in the order of $10^4/10^5$ configurations. While
optical lattice experiments might face challenges in dealing with such
statistics, other synthetic matter platforms such as Rydberg atoms in optical
tweezers and trapped ions have already achieved these regimes thanks to, in
large part, sub-Hz repetition rates. Solid state platforms are also capable of
generating such large statistics.

Another possible future direction is to apply our approach to systems that
suffer from the sign problem, to understand whether the latter reflects
intrinsically onto the dimensionality of the data
set~\cite{Westerhout_2020,Szabo2020}, or rather influences other geometrical
properties (e.g., curvature). For instance, one may start by performing
simulations in regimes where the sign problem is particularly mild (for
instance, at high temperature), and track the complexity of the data structure
as sampling becomes increasingly difficult. Another approach would be to extend
the present method to Determinantal QMC based on auxiliary
fields~\cite{Singh2017, Broecker_2016, dosSantos2003} or to adapt it to
meron-type cluster techniques~\cite{Wiese1999}, in order to allow comparisons
between different algorithms. Note that similar ideas might also be applied to
other quantum mechanical objects, such as complex-valued Wigner functions.

Finally, let us mention possible connections of our work to recent efforts to
define new classes of variational artificial-neural-network (ANN) quantum
states. Physical data sets typically lay in a manifold whose $I_d$ is lower
than the actual number of coordinates, as we extensively illustrate here.  We
believe that understanding the topography of such complex manifolds is the key
to provide a data-based comprehension of ANN quantum states' complexity. In
this sense, the $I_d$ can provide an \textit{elementary tool} for exploring the
influence of the input data structure on learning ANN quantum states
\cite{Goldt2020}. Unlike conventional variational approaches (e.g., based on
tensor-network \textit{ansatze}), where entanglement parametrizes complexity,
ANN quantum states still lack a measure of the latter. The intrinsic dimension
should be able, e.g., to give information about the number of ANN parameters
(or layers) necessary to describe a given quantum state: for the particular
case of autoencoders, we note that the intrinsic dimension is known to provide
rigorous bounds on the functioning of the network depending on the dimension of
its 'bottleneck' layer. In this context, our analysis clearly shows that the
dimensionality of the space could lead to a considerable simplification of the
data structure: this clearly points to the fact that ANN (and, more
specifically, autoencoders) could be particularly well suited to capture
quantum criticality. Furthermore, within our framework, the importance of
non-Abelian symmetries in shaping the data structure is particularly clear: not
resolving such symmetries leads to a parametrically enhanced connectivity, that
will likely affect the representative power of finite-depth ANN~\cite{Choo19}.
Going beyond these simple observations, our analysis may stimulate the
investigation of whether other features of the path integral manifold, such as
curvature or topology, are more challenging for ANN representations, and
whether those could be of use to understand the relationship between ANN and
tensor networks (see, e.g, Ref.~\cite{Glasser_2018, Pastori_2019,
Collura_2021}), both in the case of pure states, and in the case of mixed
states. Similar considerations could be extended to ANN inspired by quantum
field theory~\cite{bachtis2021quantum}.

\begin{figure*}[t]
\begin{center}
{\centering\resizebox*{7.7cm}{!}{\includegraphics*{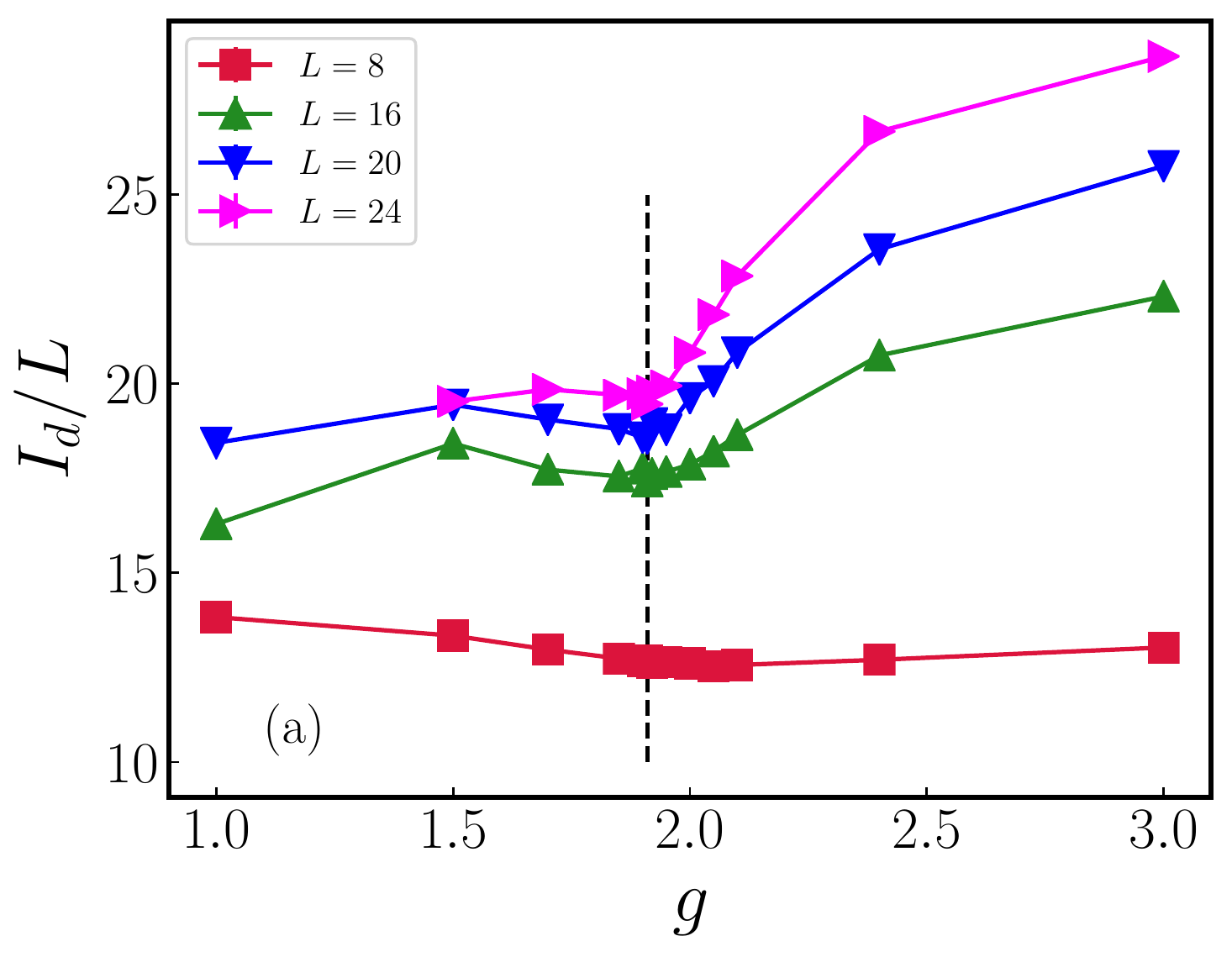}}}
{\centering\resizebox*{8.6cm}{!}{\includegraphics*{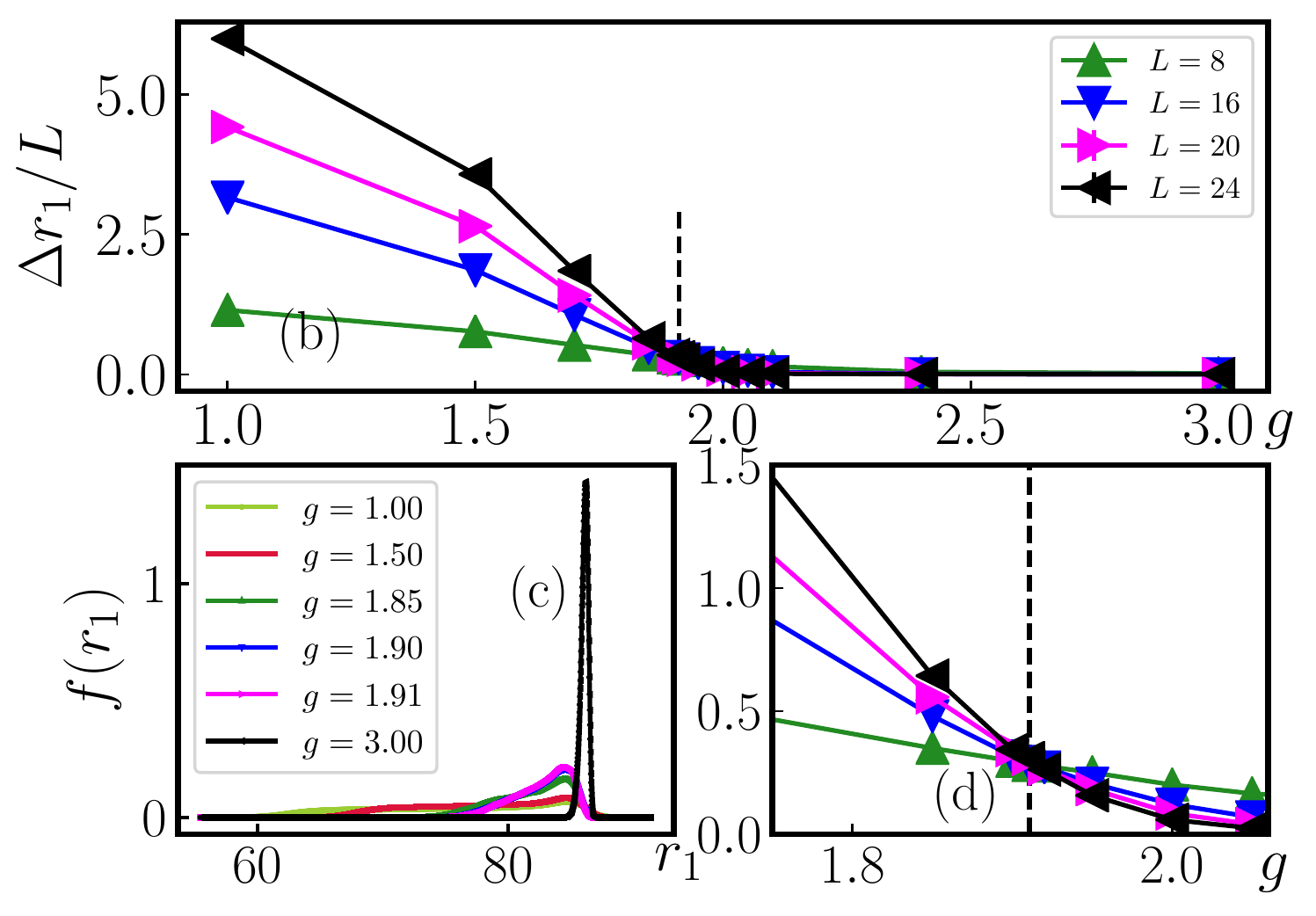}}}
\end{center}
  \caption{\textit{Heisenberg columnar dimer model.} Panel (a): $I_d$ (rescaled
  with the system size) as a function of the coupling ratio $g$. Panel (b):
  rescaled variance of the NN distance distribution function, plotted as a
  function of $g$. Panel (c): NN distance distribution functions for selected
  values of $g$. Panel (d): magnification of the curve-crossing region of panel
  (b). For all data shown here, we considered $\beta$-slices configurations,
  with $\beta = L$. In all panels, the vertical dashed line corresponds to the
  critical point $g = g_c$.}
\label{FigDi} 
\end{figure*}

\section{Acknowledgements} 

We acknowledge useful discussions with A. Browaeys, M. Heyl, A. Laio, R. T. Scalettar, and V. Vitale, and thank X. Turkeshi for collaboration on a related work. The work of AA, MD and TMS is partly supported by the ERC under grant number 758329 (AGEnTh), by the Quantera programme QTFLAG, by the MIUR Programme FARE (MEPH), and has received funding from the European Union's Horizon 2020 research and innovation programme under grant agreement No 817482. This work has been carried out within the activities of TQT. 

\appendix

\section{2D second-order transition: Heisenberg columnar dimer model}
\label{appDi}

We now provide further evidence for the conclusions drawn in Secs.
\ref{sec.2ddm} and \ref{sec:sym}. An important aspect that we investigate is if
our results are, indeed, solely determined by universal properties of the
underlying QCP or if particular features of the system (e.g., the lattice
geometry) can affect them. For example, for the bilayer geometry, one could
argue that in the regime of almost decoupled layers (i.e., $g \ll 1$),
uncorrelated DOF, in principle, may influence the behavior of data-based
quantities. To address this issue, we consider a model describing an
AFM-paramagnetic transition, but with a different lattice geometry than the
bilayer: namely, the single-layer Heisenberg columnar dimer model
\cite{Matsumoto2001, Wenzel2009}
\begin{equation}    
\label{eq.ham-di}
  H = J \sum_{\langle i,j \rangle} \boldsymbol{S}_{i} \cdot
  \boldsymbol{S}_{j} + J g \sum_{\langle i,j \rangle^{\prime}} \boldsymbol{S}_{i} \cdot
  \boldsymbol{S}_{j},
\end{equation}
where $J$ and $Jg$ are exchange couplings constants defined on two set of bonds
on a square lattice $\langle i,j \rangle$ and $\langle i,j \rangle^{\prime}$,
respectively, following the notation of ref.~\cite{Wenzel2009}. In our
simulations we set $J = 1$. The ground-state properties of this model are
equivalent to the bilayer Heisenberg model, i.e., it displays an
$SU(2)$-symmetry-broken antiferromagnetic phase and an $SU(2)$-disordered phase
for weak and strong values of $g$, respectively. QMC simulations show that the
AFM-paramagnetic quantum phase transition takes place at $g_c = 1.9096(2)$
\cite{Wenzel2009}, and belongs to the same universality class of the
three-dimensional $O(3)$ Heisenberg model.

Our simulations are performed at inverse temperature $\beta = L$, and we
consider $\beta$-slices data sets containing $N_r= 5 \times 10^4$ configurations.

In Fig. \ref{FigDi} we show the $I_d$ and the $\Delta r_1$ as function of $g$
for different system sizes $L$. Overall, our results confirm the conclusions
drawn in Sec. \ref{sec.2ddm} for the bilayer Heisenberg model. Indeed, in the
vicinity of the QCP (i) the $I_d$ features a local minimum, and (ii) $\Delta
r_1/L$ exhibits an (almost) $L$-independent behavior; the transition can then
be identified (for sufficiently large system sizes) by the crossing point of
$\Delta r_1/L$ curves for different values of $L$. These results highlight how
universal properties of the underlying QCP solely determine generic features of
data structures. Moreover, we observe that inside the $SU(2)$-symmetry-broken
antiferromagnetic phase, $\Delta r_1/L$ exhibits an extensive behavior,
confirming the important role of (broken) non-Abelian symmetries.

%%%%%%%%%%%%%%%%%%%%%%%%%%% Bibliography %%%%%%%%%%%%%%%%%%%%%%%%%%%%%%%%

\phantomsection
\addcontentsline{toc}{chapter}{Bibliography} 
\bibliography{ID}

\end{document}